\documentclass[11pt]{article}

\usepackage{jheppub} 

\usepackage[T1]{fontenc} 

\usepackage{amsmath,amssymb}

\newcommand{\pd}{\partial}
\newcommand{\e}{{\rm e}}

\newcommand{\consim}{\overset{\underset{\mathrm{c}}{}}{\sim}}

\newcommand{\D}{{\cal D}}           
\newcommand{\J}[1]{{\cal J}_{#1}}    
\newcommand{\K}[1]{{\cal K}_{#1}}    
\newcommand{\U}{{\cal C}}           
\newcommand{\ReS}{{{\rm Re} S}}     
\newcommand{\ImS}{{{\rm Im} S}}     
\newcommand{\imag}{{\rm i}}     

\title{
\boldmath
Lefschetz thimble structure in one-dimensional lattice Thirring model 
at finite density}

\author[a]{Hirotsugu Fujii,}
\author[b]{Syo Kamata}
\author[a]{and Yoshio Kikukawa}

\affiliation[a]
{Institute of Physics, University of Tokyo, Tokyo 153-8092, Japan}
\affiliation[b]
{Department of Physics, Rikkyo University, Tokyo 171-8501, Japan}

\emailAdd{hfujii@phys.c.u-tokyo.ac.jp}
\emailAdd{skamata@rikkyo.ac.jp}
\emailAdd{kikukawa@hep1.c.u-tokyo.ac.jp}

\abstract{
We investigate Lefschetz thimble structure of the complexified 
path-integration
in the one-dimensional lattice massive Thirring model with finite
chemical potential. The lattice model is formulated with staggered
fermions and a compact auxiliary vector boson (a link field), and the
whole set of the critical points (the complex saddle points) are
sorted out, where each critical point turns out to be 
in a one-to-one correspondence with a singular point of the effective action
(or a zero point of the fermion determinant). For a subset of critical
point solutions in the uniform-field subspace,
we examine the upward and downward cycles and 
the Stokes phenomenon with varying the chemical potential,
and we identify the intersection numbers to determine the thimbles
contributing to the path-integration of the partition function. 
We show that the original integration path becomes 
equivalent to a single Lefschetz thimble at small and large chemical
potentials, while in the crossover region multiple thimbles must 
contribute to the path integration.  
Finally, reducing the model to a uniform field space, 
we study  the relative importance of multi-thimble
contributions and their behavior toward continuum and low-temperature
limits quantitatively, and see how the rapid crossover behavior
is recovered by adding the multi-thimble contributions at low temperatures.
Those findings will be useful for performing Monte-Carlo simulations
on the Lefschetz thimbles.}

\begin{document} 
\maketitle
\flushbottom

\section{Introduction}



The sign problem is the longstanding obstacle which prevents us
from applying nonperturbative lattice simulations directly 
to the physical systems with complex actions, 
including quantum chromodynamics (QCD) 
at finite baryon chemical potential $\mu$.
The fermion determinant at finite $\mu$ becomes complex, which
invalidates the importance sampling algorithm. 
In contrast, the determinant is real at finite temperature ($T$) with
$\mu=0$,
and lattice simulations of QCD have proved now to be a reliable
nonperturbative method to evaluate (e.g.) the equation of state
of strongly interacting matter.
Nonetheless, studies of QCD-inspired models at finite $T$ and $\mu$
have suggested a variety of phase changes from nuclear liquid-vapor transition, 
to chiral symmetry restoration, and to color-superconducting phase
transition,~etc.  
With this situation, in order to unveil the QCD
phase diagram from the first principles,
many attempts have been made to circumvent the sign problem
in lattice QCD simulations,
although the complete resolution is still not available\cite{deForcrand:2010ys}.




To study the physical systems with complex actions,
two alternative approaches have attracted much attention recently 
-- 
complex Langevin equation\cite{Parisi:1984cs,Klauder:1983zm,Klauder:1983sp}
and Lefschetz thimble integration\cite{Witten:2010cx,Witten:2010zr,Pham:1983},
both of which involve complexification of the dynamical field variables.

Statistical sampling with the complex Langevin equation has been applied to
various models\cite{Aarts:2008rr,
Aarts:2008wh,
Aarts:2009hn, 
Aarts:2009yj,
Aarts:2009dg,
Aarts:2009uq,
Aarts:2010aq,
Aarts:2010gr,
Aarts:2011ax,
Aarts:2011sf,
Aarts:2011zn,
Seiler:2012wz,
Aarts:2012ft,
Pawlowski:2013pje,
Pawlowski:2013gag,
Aarts:2013bla,
Aarts:2013uxa,
Aarts:2013uza, 
Sexty:2013ica,
Aarts:2013fpa,
Giudice:2013eva,
Mollgaard:2013qra,
Sexty:2013fpt,
Aarts:2013nja,
Bongiovanni:2013nxa,
Aarts:2014nxa,
Aarts:2014bwa, 
Sexty:2014zya,
Sexty:2014dxa,
Bongiovanni:2014rna,
Aarts:2014kja,
Aarts:2014cua,
Aarts:2014fsa,
Mollgaard:2014mga,
Makino:2015ooa,
Aarts:2015oka, 
Nishimura:2015pba, 
Aarts:2015yba,
Nagata:2015uga,
Fodor:2015doa}, 
including the massive Thirring model with chemical potential\cite{Pawlowski:2013pje,Pawlowski:2013gag}, as testing grounds, 
and it is successful in some cases but not in other cases.
A formal proof for the correctness of the method 
has been elaborated under certain conditions\cite{Aarts:2009uq,Aarts:2011sf},
but full justification of the complex Langevin approach is not 
established, where logarithm terms such as the ferimon determinant
in the action cause a subtlety\cite{Nishimura:2015pba}.
Noteworthily, complex Langevin simulations have been applied to 
full QCD at finite $T$ and $\mu $\cite{Sexty:2013ica,
Sexty:2013fpt,Aarts:2014bwa,Sexty:2014zya,Sexty:2014dxa,
Aarts:2014kja,Aarts:2014fsa,Aarts:2015yba,Fodor:2015doa},
showing consistent results with those obtained by the reweighting method in 
the parameter region where both methods are stable\cite{Fodor:2015doa}.


Path integration on the Lefschetz thimbles was introduced in the study
of analytic property of gauge theories
\cite{Witten:2010cx}, and it was soon recognized 
as a mathematically sound way to resolve the sign problem 
\cite{Cristoforetti:2012su,Cristoforetti:2013wha,Fujii:2013sra}.
It can be regarded as a functional generalization of the
steepest descent method of complex analysis.
In this approach the original integration cycle is deformed to a
sum of the curved manifolds, called Lefschetz thimbles, 
in the complexified field space. 
On a thimble the imaginary part of the action $\ImS$ is constant, and
this property allows importance sampling with the weight $\e^{-\ReS} \ge 0$. 
This advantage was first applied to numerical simulations
for 4-dimensional $\lambda \phi^4$ theory with chemical potential with
use of Langevin \cite{Cristoforetti:2013wha}  and 
hybrid Monte Carlo (HMC)\cite{Fujii:2013sra} algorithms 
on a single thimble, and successfully reproduced the
known results including the so-called Silver Blaze behavior\cite{Cohen:2003kd}
-- complete insensitivity of the system to $\mu$ 
below a certain critical value at $T=0$.
The residual phase problem from the Jacobian due to the curvature is
mild and can be efficiently taken into account by reweighting 
for this theory\cite{Fujii:2013sra}. 
The Lefschetz thimble integration has been examined 
in other models\cite{Mukherjee:2014hsa,
DiRenzo:2015foa,
Tanizaki:2014tua,
Kanazawa:2014qma}
and has been studied from other aspects\cite{Cristoforetti:2014gsa,
Tanizaki:2014xba, 
Tanizaki:2015pua,
Cherman:2014ofa,
Behtash:2015kna, 
Tsutsui:2015tua, 
Fukushima:2015qza}
which involve the sign problem. 
This approach also shed new light on the complex Langevin
sampling method\cite{Aarts:2013fpa,Aarts:2014nxa,Tsutsui:2015tua},
and vice versa\cite{Fukushima:2015qza}.


In this paper, we study the path integration on the Lefschetz thimbles
in the (0+1) dimensional massive Thirring model at finite chemical
potential $\mu$ \cite{Pawlowski:2014ada},
in order to clarify the effects of the fermion determinant
on the structure of the thimbles contributing 
to the partition function\cite{Kanazawa:2014qma}.
The lattice model is formulated with the staggered
fermions\cite{Kogut:1974ag, Hasenfratz:1983ba}  
and a compact auxiliary vector boson (a link field). 
This model shows a crossover transition from the low to the high
density phase at finite $T$ as a function of $\mu$, and the transition
becomes first order in $T=0$ limit. 
Furthermore the exact solution of this model is available 
on the finite lattice as well as in the continuum limit, 
and therefore one can assess the validity of the approach precisely
by comparing the results with the exact ones.

The fermion determinant has zero points on the complexified field
space and actually those zeros form continuous submanifolds on which
the effective action $S$ becomes singular.
At the same time, 
the determinant brings in many critical points, each of which 
a thimble is associated to.
We classify the critical points into subsets according to the subspaces they
belong to, and identify {\it all} the critical points and thimbles in
each subspace by noting a one-to-one correspondence between a
critical point and a zero point of the determinent.
The thimbles whose critical points are located in the uniform-field subspace
are shown to dominate the integral toward the continuum limit. 
Hence we study within the uniform-field subspace  
how the set of the contributing thimbles to the partition function
changes via the Stokes phenomenon as the chemical potential $\mu$ varies.
We will see that in the crossover region 
multi-thimble contributions are inevitable,
and become more significant for small inverse coupling
and/or in low temperature limit. We study this interplay in more detail
by reducing the model degrees of freedom to the uniform-field subspace
and show how the crossover behavior is reproduced
as adding the multi-thimble contributions to the observables.


This paper is organized as follows.
In section~\ref{sec:LatticeThirringModel}
we introduce the (0+1) dimensional massive Thirring model with
chemical potential on the lattice in terms of the staggered fermions and the
compact auxiliary vector boson.
In section~\ref{sec:CP&Zeros},
after a briefly review of the Lefschetz thimble approach,
we study the critical points and determinant zeros of the lattice
model in the complexified field space. The critical points
are classified by the subspaces they live,  
and all the critical points are identified.
In section~\ref{sec:ThimbleMu0},
we study the thimble structure of the model at $\mu=0$,
discuss the importance of each thimble by looking 
at the relative weight at $\mu=0$.
In section~\ref{sec:Stokes}, we show within the
uniform-field subspace the change of the thimble
structure with increasing the chemical potential $\mu$ via the Stokes
phenomenon, and show that the multiple thimbles contribute to the
partition function in the crossover region.
In section~\ref{sec:ModelStudy}, taking the uniform-field subspace, we
examine the validity of the single thimble approximation, 
and discuss the continuum and low temperature limits. Especially in
the low temperature limit, the importance of the multi-thimble
contributions are clarified. 
Section~\ref{sec:Summary} is devoted to summary and discussions.
The exact solution of the model is derived in Appendix~\ref{app:exact}.

\section{one-dimensional massive Thirring model on the lattice
\label{sec:LatticeThirringModel}}

The (0+1)-dimensional lattice Thirring model we consider in this paper 
is defined by the following action\cite{Pawlowski:2013pje,Pawlowski:2013gag,Pawlowski:2014ada}, 
\begin{align}
S_0 =& 
\beta  \sum_{n=1}^{L}  \big(1- \cos A_n \big)  \nonumber\\
& - \sum_{n=1}^{L} \sum_{f=1}^{N_f}
\bar \chi^f_n  \left\{
 \e^{i A_n +\mu a } \, \chi^f_{n+1} -  \e^{-i A_{n-1}-\mu a } \,  \chi^f_{n-1}  
 + m  a  \, \chi^f_n \right\}
\, ,
\end{align}
where $\beta = (2 g^2 a)^{-1}$,  $m a$, $\mu a$ 
are the inverse coupling, mass and chemical potential in the lattice unit, 
and  $L$ is the lattice size which defines the inverse temperature as
$1/T \equiv La$.
The  fermion field has $N_f$ flavors and
satisfies the anti-periodic boundary conditions: 
$\chi^f_{L+1}=-\chi^f_1$, $\chi^f_0=-\chi^f_L$, 
$\bar \chi^f_{L+1}=-\bar\chi^f_1$, and  $\bar\chi^f_0=-\bar\chi^f_L$. 
The partition function of this lattice model is defined by the path-integration,
\begin{align}
\label{eq:path-integral-original}
Z =& \int \D A  \D  \chi \D  \bar\chi \; \e^{- S_0} 
\nonumber\\
  =&  \int_{-\pi}^{\pi} \prod_{n=1}^L d A_n  \, 
     \e^{- \beta  \sum_{n=1}^{L}  \big(1- \cos A_n \big) } 
   \, (\det D \, [A])^{N_f}, 
\end{align}
where $D$ denotes the lattice Dirac operator, 
\begin{align}
( D \chi )_n =
 {\rm e}^{i A_n +\mu a } \, \chi^f_{n+1} -  {\rm e}^{-i A_{n-1}-\mu a} \,  \chi^f_{n-1}  
 + m a \, \chi^f_n . 
 \end{align}
The functional determinant of $D$ can be evaluated explicitly
(see Appendix \ref{app:exact} for derivation) as 
\begin{align}
\label{eq:detD-A}
 \det D \, [A] = \frac{1}{2^{L-1}} 
\Big[ 
\cosh (L  \hat \mu + i {\scriptstyle \sum_{n=1}^L } A_n) + \cosh L \hat m 
\Big]  
\end{align}
with $\hat \mu = \mu a$, $\hat m=\sinh^{-1} m a$.
It is not real-positive for $\mu \not = 0$ in general, but
instead it has the property 
 $\left( \det D[A]\vert_{+\mu} \right)^\ast 
= \det D[-A]\vert_{+\mu} = \det D[A]\vert_{-\mu}$. 
This fact can cause the sign problem in Monte Carlo simulations.

\begin{figure}[tb]
\begin{center}
\includegraphics[width=0.45\textwidth]{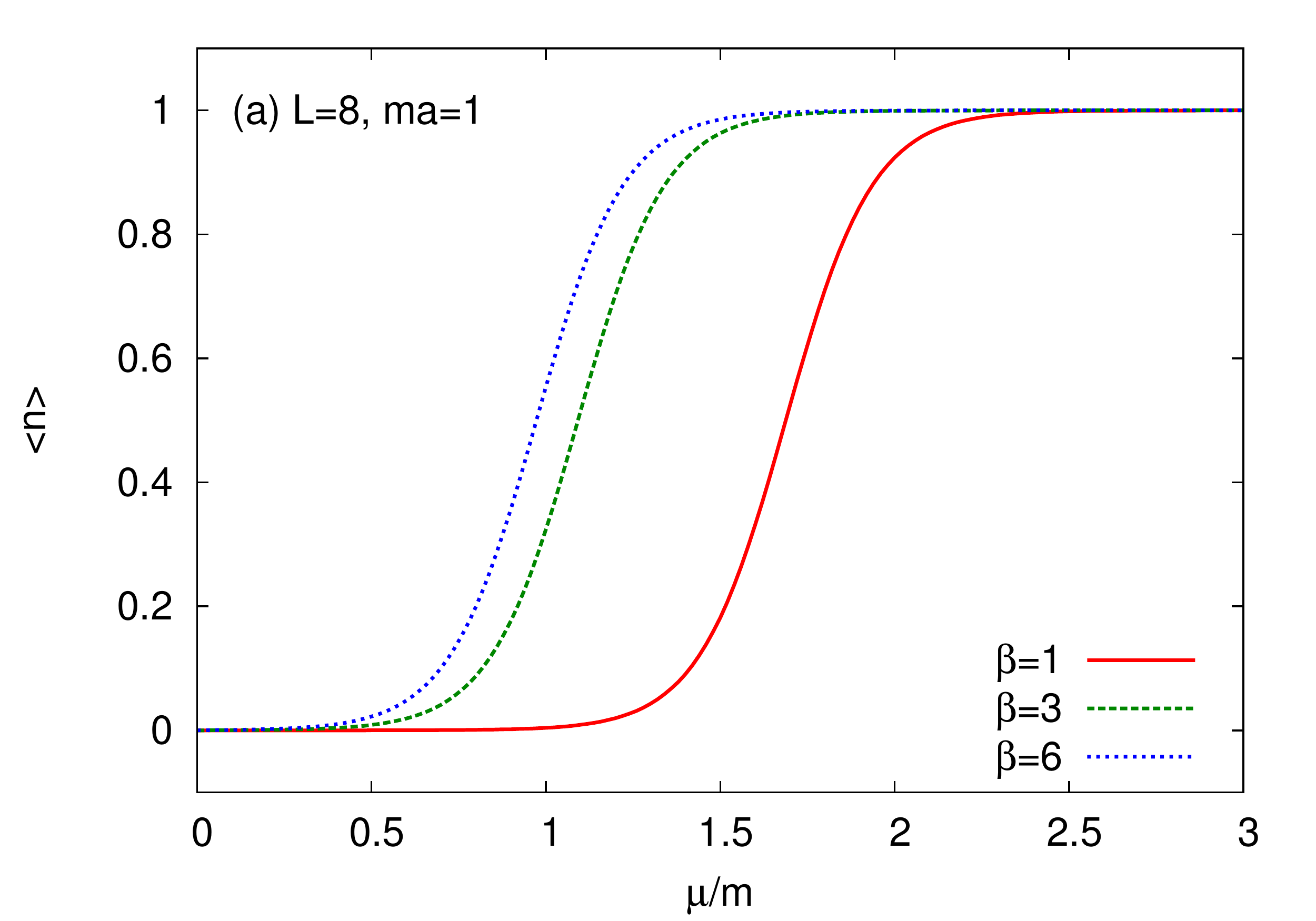}
\hfil
\includegraphics[width=0.45\textwidth]{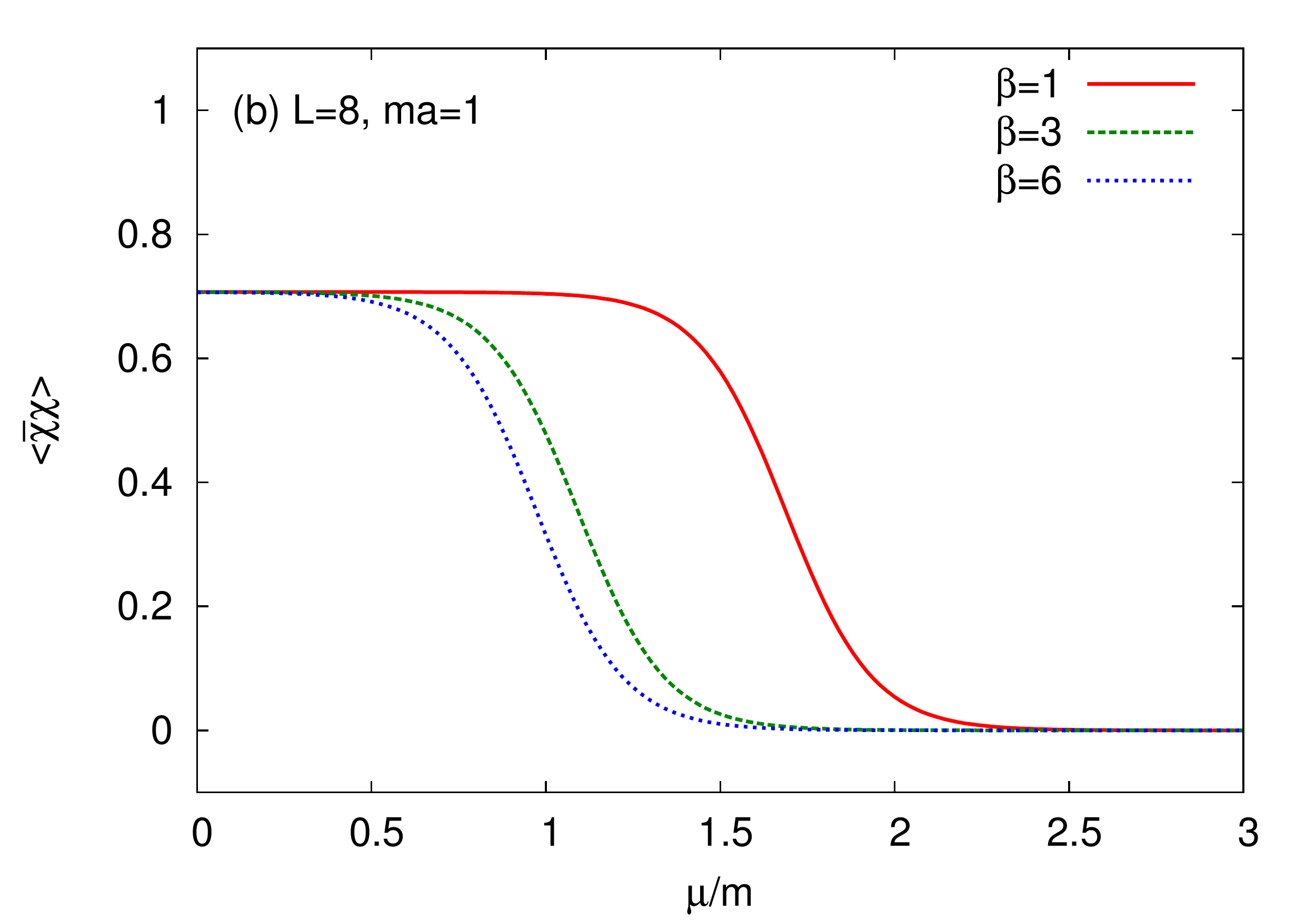}
\end{center}
\caption{
(a) Fermion number density and (b) scalar condensate with $L=8$,
$ma=1$ for $\beta =1$ (solid), 3 (dashed), and 6 (dotted).
}
\label{fig:n-c-exact-L=8-m=1-beta=1-3-6} 
\end{figure}

This lattice model is exactly solvable in the following sense. 
The path-integration over the field $A_n$ can be done 
explicitly and the exact expression of the partition function is
obtained ($N_f=1$) as 
\begin{align}
Z= \frac{\e^{-\beta L} }{2^{L-1}} \, 
\Big[ I_1(\beta)^L  \cosh L \hat \mu +  I_0(\beta)^L \cosh L \hat m 
\Big]
\, ,
\label{eq:Zexact}
\end{align}
where $I_{0,1}(\beta)$ are the modified Bessel functions of the first kind.
The number density and scalar condensate of the fermion field are
then obtained as
\begin{align}
\langle n \rangle 
&\equiv \frac{1}{La } \frac{\partial \ln Z}{\partial \mu}
 \nonumber\\
&= \frac{ I_{1}(\beta)^{L} \sinh L \hat \mu }
        {I_{1}(\beta)^{L} \cosh L\hat \mu + I_{0}(\beta)^{L} \cosh L
          \hat{m}}
\, ,\\
&\nonumber\\
\langle \bar{\chi} \chi \rangle
&\equiv \frac{1}{La} \frac{\partial \ln Z}{\partial m} 
\nonumber\\
&=    \frac{1}{\cosh \hat{m}}\,
\frac{ I_{0}(\beta)^{L} \sinh L \hat{m}}
     { I_{1}(\beta)^{L} \cosh L \hat \mu + I_{0}(\beta)^{L} \cosh L \hat{m} }
\, .
\end{align}
The $\mu$-dependence of these quantities are shown in 
Fig.~\ref{fig:n-c-exact-L=8-m=1-beta=1-3-6} 
for $L=8$, $ma=1$, and $\beta=1,3$, and 6.
It shows a crossover behavior in the chemical potential $\hat \mu$ (in
the lattice unit) around 
$\hat \mu \simeq \hat m + \ln (I_0(\beta)/I_1(\beta))$.

\begin{figure}[tb]
\begin{center}
\includegraphics[width=0.3\textwidth,bb=0 100 259 292]{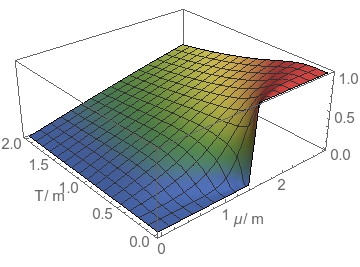}
\end{center}
\vspace*{10mm}
\caption{
Number density $\left < n \right >$ on the $T$-$\mu$ plane
in the continuum limit ($g^2/m=1/2$).
}
\label{fig:phase-structure-in-T-mu-continuum-limit}
\end{figure}

The continuum limit ($a \rightarrow 0$) of this lattice model  at finite
$T$ may be defined as
\begin{align}
\beta = \frac{1}{2 g^2 a} 
\rightarrow \infty, \qquad
L = \frac{1}{Ta}  
\rightarrow \infty \qquad
\text{with  } \, \,  
\beta / L = T/(2 g^2) \, \, \text{ fixed}.
\end{align}
In this limit the partition function scales as 
\begin{align}
Z \to \frac{1}{2^{L-1}}\left(\frac{1}{2\pi\beta}\right)^{L/2} 
\e^{\frac{3 g^2}{4T}}  
\left( \cosh \frac{\mu}{T}  + \e^{\frac{g^2}{T}}  \cosh \frac{m}{T} \right) , 
\end{align}
and the continuum limits 
of $\langle n \rangle$ and $\langle \bar \chi \chi \rangle$
are obtained as follows:
\begin{eqnarray}
\lim_{a \rightarrow 0} \, \langle n \rangle 
&=& \frac{\sinh \frac{\mu}{T} }
{ \cosh \frac{\mu}{T}+ {\rm e}^{\frac{g^2}{T}} \cosh \frac{m}{T} } , 
\nonumber\\
\lim_{a \rightarrow 0} \, \langle \bar \chi \chi \rangle 
&=& \frac{ \e^{\frac{g^2}{T}} \sinh \frac{m}{T}}
         { \cosh \frac{\mu}{T} + \e^{\frac{g^2}{T}}  \cosh \frac{m}{T} } . 
\end{eqnarray}
From these results, 
one sees that the model shows a crossover behavior in the chemical potential $\mu$ 
at non-zero temperatures $T > 0$, 
while in the zero temperature $T=0$ limit 
it shows a first-order transition at the critical chemical potential 
$|\mu_c|  = m + g^2$.  
See Fig.~\ref{fig:phase-structure-in-T-mu-continuum-limit}.

\section{Lefschetz thimble approach\label{sec:CP&Zeros}}
\subsection{Preliminaries}


Now we consider the complexification of the Thirring model on the
lattic and reformulate 
the defining path-integration of Eq.~(\ref{eq:path-integral-original}) 
by the integration over Lefschetz thimbles.
In the complexification,  
the field variables $A_n $ are extended to 
complex variables $z_n$ $(\in \mathbb{C}^L)$ and 
the action is extended to a holomorphic function given by 
$S[z] = \beta \sum_{n=1}^L (1- \cos z_n)  - \ln \det D [z]$.%
\footnote{The logarithm has branch cuts, but it does not affect the
  gradient flows as discussed below.}
For each critical point $z=\sigma$ given by the stationary condition,
 \begin{align}
 \label{eq:stationary-condition}
\left .  \frac{ \partial S [ z] } {\partial z_n} \right |_{z=\sigma}
 =0  \quad (n=1,\cdots,L), 
\end{align}
the thimble $\mathcal J_\sigma$ is defined as a union of all 
the (downward) gradient flow curves determined by
\begin{eqnarray}
\label{eq:downward-flow-equation}
\frac{d}{dt} z_n(t) =  \frac{ \partial \bar S [ \bar z] }{\partial \bar z_n} 
\quad  
( t \in \mathbb{R} )  
\qquad \text{s.t.} \quad
z(-\infty) = \sigma. 
\end{eqnarray}
The thimble so defined is an $L$-dimensional real submanifold in $\mathbb{C}^L$.
Then, according to Picard-Lefschetz theory (complexified Morse
theory),  the original path-integration region  
${\mathcal C}_\mathbb{R} \equiv [ -\pi , \pi]^L$
can be replaced with a set of 
Lefschetz thimbles%
\footnote{
We will extend this original integration region to
${\mathcal C}_\mathbb{R} \equiv ([ - \pi+i \infty, -\pi] \oplus  [
    -\pi , \pi] \oplus [\pi, \pi+i\infty])^L $
as the well-defined integration cycle.
The value of the integral is unchanged by this extension thanks to
$2\pi$ periodicity of $S$.
},
\begin{align}
{\mathcal C}_\mathbb{R}
= \sum_{\sigma}  n_\sigma {\cal J}_\sigma , 
\end{align}
where 
$n_\sigma$ stands for the intersection number between 
${\mathcal C}_\mathbb{R}$ and the dual submanifold ${\cal K}_\sigma$, 
which is another $L$-dimensional real submanifold associated to 
the same critical point $\sigma$ and  
is defined as a union of all the gradient flow curves
s.t.~$z(+\infty) = \sigma$.
With denoting the set of the critical points
as $\Sigma \equiv \{\sigma\}$,
the partition function and the correlation functions of the
lattice model can be expressed by the formulas%
\footnote{The notations for the expectation values on thimbles
here are modified from those in Ref.~\cite{Fujii:2013sra} 
for later convenience.} 
\begin{eqnarray}
\label{eq:partion-function-by-thimbles}
Z &=& \sum_{\sigma \in \Sigma}  n_{\sigma}  
\, Z_{\sigma}, 
\, \qquad
Z_{\sigma} \equiv
\int_{{\cal J}_\sigma} {\cal D}[z] \, 
{\rm e}^{-S[z]},  \\
\langle O[z] \rangle &=& 
\frac{1}{Z} \sum_{\sigma \in \Sigma}  n_{\sigma}  \, \langle O[z] \rangle_{\sigma} , 
\quad
\langle O[z] \rangle_{\sigma}
\equiv
\int_{{\cal J}_\sigma} {\cal D}[z] \, {\rm e}^{-S[z]} O[z] .
\end{eqnarray}
The functional measure $\D[z]$ along the thimble $\J{\sigma}$
is specified as 
$ \left. d^Lz \right |_{\J{\sigma}} = d^L (\delta \xi) \det U_z$
by the orthonormal basis of tangent vectors 
$\{ U_z^\alpha  \vert (\alpha=1,\cdots, L) \}$
which span the tangent space as
$ \delta z = U_z^\alpha \delta \xi^\alpha $ 
$( \delta z \in \mathbb{C}^L, \delta \xi \in \mathbb{R}^L )$.

The integration on each Lefschetz thimble is convergent because
the real part of the action increases monotonically to $\infty$
while
the imaginary part stays constant along the downward flow,
\begin{align}
& \frac{ d\, \ReS}{dt} \ge 0, \qquad \frac{ d\, \ImS}{dt} = 0
.
\label{eq:dSdt_re_im}
\end{align}
The sign problem remains in the Lefschetz thimble approach in two
facts.
First, it seems that 
when we factor out the {\it  complex weight} $\e^{-S[\sigma]}$,
the integrand of each thimble, $\e^{-(S[z]-S[\sigma])} > 0$, is real positive.
But a complex phase appears from the Jacobian factor $\det U_z$ 
in the integration, which is called residual sign problem. 
For $\lambda \phi^4$ theory it is demonstrated that 
the residual sign problem can be treated by the reweighting
method safely\cite{Fujii:2013sra}.
Second, the terms $Z_\sigma$ and $\langle O[z] \rangle_{\sigma}$ 
are actually complex quantities although the total averages
$Z$ and $\langle O[z] \rangle$ should be real.
If there is a certain symmetry in the thimble structure of the
system, one can show the cancellation of the phases 
in the sum\cite{Tanizaki:2015pua}.  
The multi-thimble contributions to the partition function
and observables will be more elaborated in this paper.

\subsection{Critical points and determinant zero points}

Given the above mathematical results, however, 
it is not straightforward to work out 
for general fermionic models all the critical points 
$\Sigma=\{\sigma\}$, the thimbles $\{ \J{\sigma} \}$,
and their intersection numbers $\{ n_\sigma\}$.
Fortunately in our lattice model, we can find
all the critical points determined by the stationary condition
Eq.~(\ref{eq:stationary-condition}).

The critical point condition for the Thirring model is written as 
\begin{align}
\frac{\pd S}{\pd z_{n}}  = 
\beta \sin z_n - 
\frac{i \sinh  (L \hat \mu + i s ) }
{   \cosh  ( L \hat \mu + i s  ) 
+  \cosh L\hat m }
=0
\qquad
\text{with } s \equiv \sum_{\ell=0}^{L-1} z_{\ell}.
\label{eq:crit}
\end{align}
The key observation is that the second term
depends on the field configuration
only through the sum $s$,
so that all $\sin z_n$ ($n=0,\cdots, L-1$) of a critical point $\sigma$
must have the same value to cancel the common second term.
Let us denote it as $\sin z$,
then the field components can be
either $z_n = z$ or $\pi - z$ and the sum $s$ is written as
\begin{align}
s= n_+ z +n_- (\pi - z) 
 = (L-2n_-) z + n_- \pi
,
\end{align}
where $n_\pm$ are the numbers of $z$ and $\pi - z$ in the components
$\{z_n\}$ with $n_+ + n_-=L$.
The critical point condition for $z$ is now explicitly written
as
\begin{align}
\beta \sin z  
-
\frac{ i \sinh \, [L \hat \mu + i (L-2n_-) z] \ }
{ \cosh \, [ L \hat \mu + i (L-2n_-) z ] \ +  (-)^{n_-}\cosh L\hat m  }
=0.
\label{eq:critcond2}
\end{align}
This can be regarded
as the critical point condition for a one-variable model;
\begin{align}
S_{n_-}
&= 
(L-2n_-)\beta (1-\cos  z  )
- \log \Big (
 \cosh \, [ L \hat \mu + i (L-2n_-) z ] \ +  (-)^{n_-}\cosh L\hat m  
\Big )
\, .
\label{eq:onevar_model}
\end{align}
The case of $n_-=0$ corresponds to a uniform field configuration,
where $z_n=z$ ($n=0, \cdots, L-1$), 
and the case $n_-=1$ means that there is one flipped component
$\pi-z, \cdots$, etc.
In the case of $n_-=L/2$, the second term of (\ref{eq:critcond2}) 
becomes independent of $z$. 
The case of $n_->L/2$ gives the same critical points
as in the case $L-n_-$ with $z \leftrightarrow \pi-z$.  
Hence we need to consider $n_-=0, \cdots, L/2-1$.

Thus we have classified the critical points with index $n_-$.
By solving the condition
Eq.~(\ref{eq:critcond2}) of the one-variable model for each $n_-$,
we can locate all the critical points of the model. 
Note that a critical point is 
{\small $\left ( \begin{array}{c} L \\ n_-\end{array}\right )$}-ply degenerated
for $n_- \ne 0$  due to the combination about which components to be flipped.

One of the distinctive features of fermionic theories is the fact that
the fermion determinant has many zero points within a compact domain
in the complexified space.
The real part of the effective action $\ReS$ 
diverges at these zeros, and therefore 
the downward cycle $\J{\sigma}$ 
may flow into one of these zeros, otherwise it must extend outward 
to the safe exterior region where ${\rm Re} S=+\infty$. 
Hence, in addition to the critical points,
we need to locate all the zeros of the fermion determinant 
\begin{align}
\det D[z] = 0 \, .
\label{eq:sing}
\end{align}

Thanks to the 
concise expression of $\det D[z]$ in Eq.~(\ref{eq:detD-A}), 
one can easily find {\it all} zero points:
\begin{align}
s_{\rm zero} = \imag L( \hat \mu \pm \hat m) + (2n+1)\pi  \qquad (n\in {\mathbb Z}).
\label{eq:szero}
\end{align}
This only fixes $s=\sum_{\ell=0}^{L-1}z_\ell$,
and thus defines submanifolds with the complex dimension $L-1$, 
embedded in the $L$ dimensional complexified configuration space.
Note that these zero points are independent of $\beta$, and 
that nonzero $\hat \mu$ simply shifts the zero points along the imaginary axis.
Restricting this submanifold of the zeros in the subspace $n_-=0$,
where $s=Lz$,
we find $2L$ isolated zeros of
\begin{align}
z_{\rm zero}= \imag (\hat \mu \pm \hat m) 
+ \frac{2n+1}{L}\pi  \qquad (n\in {\mathbb Z} {\rm ~mod~} L),
\label{eq:zzero0}
\end{align}
while in the subspace ($n_-=1$) with a single link flipped to $\pi-z$
(and thus $s=(L-2n_-)z +\pi$), 
we have  $2(L-2)$ zeros of
\begin{align}
z_{\rm zero}= \imag  \frac{L}{L-2} (\hat \mu \pm \hat m) 
+ \frac{2n}{L-2}\pi  \qquad (n\in {\mathbb Z} {\rm ~mod~} L-2).
\label{eq:zzero1}
\end{align}

\begin{figure}[tb]
\begin{center} 
\includegraphics[width=41mm,bb=80 0 330 257]{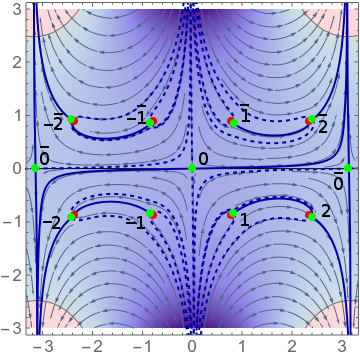}
\hspace{5mm}
\includegraphics[width=42mm,bb=0 0 259 257]{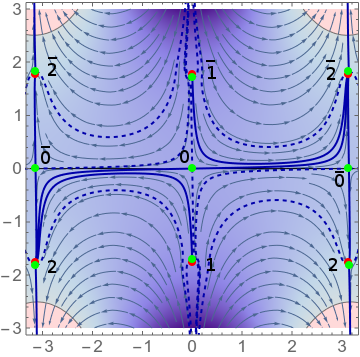}
\caption{
Critical points (green dots) and determinant zeros (red dots)
of the Thirring model 
with $L=4$ and $ma=1$
at $\hat \mu=0$ within the subspaces of $n_-=0$ (left) and 1 (right).
We set $\beta = 3 - 0.1 \imag $. 
Gradient flows are drawn with arrows.
We assign numbers to the critical points as
$\sigma_{i, \bar i}$ here.
The downward $\J{\sigma}$ and upward  $\K{\sigma}$ 
cycles of a critical point $\sigma$ are shown 
with solid and dashed lines, respectively.
The brighter background indicates the larger value of $\ReS$
(in arbitrary unit).}
\label{fig:reg_beta} 
\end{center}
\end{figure}

Figure~\ref{fig:reg_beta} shows 
two sections of the gradient flows 
in the uniform-field subspace ($n_-=0$; left) and
in the subspace with one link flipped ($n_-=1$; right)
of the model with $L=4$, $\beta=3-0.1 \imag$ and $ma=1$ 
at $\hat \mu=0$. (The reason for complex $\beta$ will be explained
in the next section.) 
Globally, the flows are streaming out of the remote points
$z=\pm  \imag \infty$ and flowing away towards 
the safe remote points $z=\pm \pi  \pm \imag \infty$.
We solve Eq.~(\ref{eq:critcond2}) numerically,
and find ten (eight) critical points%
\footnote{
The two critical points located at $z=\pm \pi$ are identical.
}
for $n_-=0$ (1), 
as shown with green dots in Fig.~\ref{fig:reg_beta}.
For later convenience, we have numbered the critical points as shown here.
We also put the zero points of the determinant $\det D[z]$
with red dots.
Each critical point apparently pairs up with a zero point next to it,
besides the two sitting at the origin and~$\pm \pi$.

Now that we have identified all the critical points and the zeros of the
Thirring model, we can study the structure of
the Lefschetz thimbles of the model in detail.

\section{Thimble structure at $\mu=0$
\label{sec:ThimbleMu0}}
\subsection{Bosonic theory}

\begin{figure}[tb]
\begin{center}
\includegraphics[width=0.3\textwidth,bb=0 100 259 357]{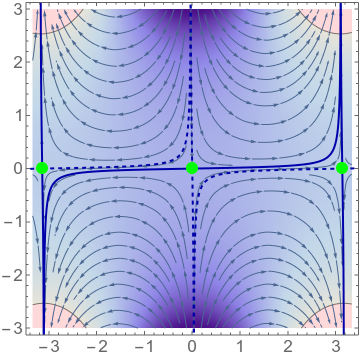}
\end{center}
\vspace*{10mm}
\caption{
Downward flow, critical points of free theory for $\beta = 3 - 0.1 \imag $
in the complex $z$ plane. 
The horizontal (vertical) axis is for the real (imaginary) part.
\label{fig:GradFlowFree}
}
\end{figure}

It would be instructive to start our discussion
with the bosonic theory without fermions, 
$S[z] \equiv \sum_{n=0}^{L-1} S_n(z_n) = 
\sum_{n=0}^{L-1}\beta (1-\cos z_n)$,
whose complexified configuration space is a direct product of
$(S^1\times {\mathbb R})^L$.
The  downward flow is simply given by
\begin{align}
\frac{dz_n}{dt}={\frac{\pd \bar S[ \bar z]}{\pd \bar z_{n}}}  = 
\bar \beta \sin \bar z_n 
\; ,
\label{eq:gradflowfree}
\end{align}
which is depicted in Fig.~\ref{fig:GradFlowFree} 
for a certain $z_n$ with $\beta=3 - 0.1 \imag$.

Let us focus on this complex $z_n$ plane for a moment.
The action is periodic in the direction of the real axis, so that the
configuration space is equivalent to $S^1 \times {\mathbb R}$, a cylinder.
There are two critical points, $z_n=0$ and $\pm \pi$ (shown in green dots),
\footnote{Note that $z_n=\pm \pi$ are the same point
on $S^1 \times {\mathbb R}$.}
corresponding to the Gaussian and doubler solutions, respectively.
The downward cycle (thimble) $\J{0}$ associated to $z_n=0$,
 extends to the ``safe'' exterior regions toward
$z_n= \pm \pi \pm \imag \infty $ depicted with light-red color at the corners.
There is another thimble $\J{- \pi}$
associated to the doubler solution $z_n= - \pi$,
which connects these two ``safe'' regions vertically 
along the imaginary direction.
In other words, the two safe regions are connected by 
two cycles with and without winding around the cylinder. 
These cycles constitute the base of homology of 
this restricted space $S^1 \times {\mathbb R}$.

We notice here that the original integration path
from $z_n=-\pi$ to $\pi$ is ill-defined as a homology cycle.
A well-defined downward cycle should extend to a ``safe'' region
where the Morse function ($h=-\ReS$) approaches
$-\infty$\cite{Witten:2010cx}.  
Actually, the thimble $\J{0}$ coincides with this original path
only for real $\beta$, which is the very parameter 
for the Stokes phenomenon to occur between $z_n=0$ and $\pm \pi$
(the action is real at both points; $S_n=0$ and $2\beta$).
Hence in Fig.~\ref{fig:GradFlowFree} 
we have added nonzero imaginary part to the coupling $\beta = 3 -0.1 \imag$
\footnote{If we take $\beta = 3 +0.1 \imag$, 
the flow structure is just reflected
about the imaginary axis from Fig.~\ref{fig:GradFlowFree}.}
to make the thimble $\J{0}$ well-defined.

Thanks to the periodicity of the action $S_n(z)$, 
we can exptend the original integration path 
without changing the value of $Z$
to a U-shaped integration cycle 
which starts at $z_n=-\pi+\imag \infty$ 
and comes down along the imaginary direction to $z_n=-\pi$ 
then moves along the real axis to $z_n=\pi$, 
and goes up to $z_n=\pi+ \imag \infty$
\footnote{One may choose alternatively
the cycle which connects $z_n=\pm \pi -\imag \infty$
passing through $z_n=0$, which does not change the discussions
below.}.
This U-shaped cycle 
(which we simply denote with $\U$) 
is equivalent to the sum of the two thimbles:
\begin{align}
\U \sim \J{0} + \J{-\pi}
\; .
\end{align}
Here we set the orientation of the thimbles 
so that ``+'' sign is appropriate here. 
One can confirm that both the upward cycles 
$\K{0}$ and $\K{-\pi}$ intersect
this integration cycle $\U$.


There are $2^L$ critical points 
in the (0+1) dimensional bosonic theory with $L$ lattice sites
from combinatorics,
and its thimble structure is obtained as a direct product of
the thimbles $\J{0}$ and $\J{-\pi}$.
The integration cycle equivalent to the original integration path
is symbolically written as
\begin{align}
\U^L \sim (\J{0} + \J{-\pi})^L
\; .
\end{align}
The safe  exterior region where the real part of
the action $\ReS$ diverges
has the complex dimension $(L-1)$ because it is 
characterized by the condition
$\sum_{n=0}^{L-1}(1- \cos\, z_n) =  \infty$,
{\it i.e.,} at least one of $\{z_n\}$ is fixed to $\pi\pm \imag \infty$.

We comment on the continuum limit ($\beta \to  \infty$ with fixed $\beta/L$). 
In this limit
the contribution to the partition function 
from each variable becomes Gaussian:
\begin{align}
\int_{-\pi}^{\pi} \frac{dz}{2\pi} e^{-\beta (1-\cos z)} = I_0(\beta) e^{-\beta}
\to \frac{1}{\sqrt{2\pi \beta}}
.
\end{align}
The integration along the vertical path, which we have added
to make the integration cycle well-defined,
becomes irrelevant giving only a contribution which is exponentially suppressed.
For example,
\begin{align}
\int_{\pi+ \imag \infty}^{\pi- \imag \infty} \frac{dz}{2\pi} \e^{-\beta (1-\cos z)} \to
- \frac{\imag }{\sqrt{2\pi \beta}} e^{-2 \beta}
.
\end{align}
Thus we see that the doubler contribution $\J{-\pi}$ is
suppressed by $\e^{-2\beta}$ and the free theory with $L$ degrees of freedom
is correctly reproduced by the integration on the thimble $\J{0}^L$.

\subsection{Thirring model
\label{subsec:thimble}}

We have already idintified the critical points and determinant
zeros of the Thirring model in the previous section
and shown them in Fig.~\ref{fig:reg_beta}.
There we also noticed a certain correlation between a critical
point and a determinant  zero. Now let us look at the thimble
structure  of the model at $\hat \mu=0$.

In the uniform-field ($n_-=0$) subspace shown in
Fig.~\ref{fig:reg_beta} (left), 
the thimble ${\cal J}_{\sigma_0}$ extends from one 
safe remote $z=-\pi - \imag \infty$ to another safe remote 
$z= \pi+ \imag \infty $ passing through 
the critical point $\sigma_0$
at the origin,
and the U-shaped cycle is still
equivalent to the sum of the two thimbles,
associated to the Gaussian and doubler critical points:
\begin{align}
\U \consim \J{\sigma_0} + \J{\sigma_{ \bar 0}}
\; ,
\end{align}
where ``$\consim$'' indicates the equivalence
as the integration cycles under the constraint
$n_-=0$.
In the subspace of $n_-=1$ (Fig.~\ref{fig:reg_beta} (right)),
on the other hand,
the critical point $\sigma_0$ contains one doubler component
$\pi-z$,
and the thimble ${\cal J}_{\sigma_0}$ ends at determinant 
zeros%
\footnote{
Note that we use the same notation $\sigma_0$ 
for the critical points in $n_-=0,1$ subspaces without any confusion.}.
The U-shaped cycle within $n_-=1$ space
is covered by the sum of four thimbles:
\begin{align}
\U \consim 
 \J{\sigma_{\bar 2}} + \J{\sigma_0} + \J{\sigma_{ \bar 0}}
 - \J{\sigma_{\bar 2}}
\; ,
\end{align}
with two $\J{\sigma_{\bar 2}}$ contributions canceling out in the end.

The strong correlation between a critical point and a zero point
may be expected by noticing the fact that 
because a zero point $z_{\rm  zero}$ is a simple pole of the flow field,
one can always find in its vicinity the point on which the first term 
of Eq.~(\ref{eq:critcond2}) can be counter-balanced 
by the would-be pole contribution,  especially when $\beta$ is large.

One can also understand the paring between them
by considering the thimble structure of the one-variable
model assigned by $n_-$, where 
a thimble $\J{\sigma}$ becomes a line segment 
associated to a critical point $\sigma$
and connects between the zeros and/or safe remote
points $z=\pm \pi \pm \imag \infty $.
In $n_-=0$ case, for example, 
two safe remote points are connected by the two thimbles,
$\J{\sigma_0}$ and $\J{\sigma_{\bar 0}}$.
Because the thimbles form the basis of independent cycles,
no trivial loops are allowed.
That is, a set of thimbles must be a connected skeleton graph
on $S^1 \times {\mathbb R}$ subspace assigned by $n_-$.
One can add a new critical point, which is 
accompanied by a new thimble, only when one has a new zero point.
Hence 
the number of thimbles coincides with the number of the critical points,
and furthermore with the number of the end-point zeros 
(including two safe remote points)
in our model.

The formulas (\ref{eq:zzero0}) and (\ref{eq:zzero1})
with $L=4$ give us 
eight and four zeros for $n_-=0$ and 1, respectively, as seen
in Fig.~\ref{fig:reg_beta}.
Adding two remote zeros, we have ten thimbles, ten critical points,
and ten zeros for $n_-=0$,
and six of those for $n_-=1$. We have just two thimbles
in $n_-=2$ subspace because we have no determinent zeros there.


\begin{table}
\begin{center}
\begin{tabular}{cccccccccc}
\hline
$L$ &$n_-$ & $0$ & $1$ & $2$ & $3$ & $4$ & $\cdots$ & $\bar 0$&$2\beta (L-2n_-)$  \\
\hline
4 & 0  & $-$0.8  &  1.6 &25.5 & --- & --- &  ---  & 23.2 & 24  \\
  &  1  &  5.3   &$-$3.6 &32.4 & --- & --- & ---   & 17.3 & 12  \\
\hline
8 & 0  & 1.7  & 2.5  & 32.0 & 73.1  & 101.5 & --- & 97.7 & 96   \\
  &  1  & 13.7 & 8.5  & 30.8 & 74.4 & 95.7 &  ---  & 85.7 & 72   \\
\hline
16& 0  & 7.1  & 7.5  & 37.6 & 92.5 & 164  &$\cdots$  & 391  &  384  \\
  &  1  & 31.2   & 26.7 & 44.6 & 93.5 & 164  &$\cdots$ & 367  &  336  \\
\hline
\end{tabular}
\caption{$\ReS$ at the critical points $\sigma_i$ ($i=0,1,2,3,4,\bar 0$)
with $L=4 k$, $\beta=3 k$, and $ma=1/k$ ($k=1,2,4$)
at $\mu=0$. The rightmost column shows the difference of $\ReS$
between $\sigma_0$ and $\sigma_{\bar 0}$.
}
\label{tab:ReS}
\end{center}
\end{table}

Note that a thimble $\J{\sigma}$ is not a simple curve
but extends with real dimension $L$,
and its section with the subspace is seen as a curve 
in Fig.~\ref{fig:reg_beta}.
For example, 
integration on the thimble $\J{\sigma_0}$ associated to the Gaussian 
critical point $z=\sigma_0$ in $n_-=0$ subspace contains 
the perturbative fluctuations in all the directions 
around $z=\sigma_0$.

Finally in this subsection,
let us look at the real part of the action $\ReS[\sigma_i]$
at these critical points for real $\beta=3$, which is
listed in the first row ($L=4$) of Table~\ref{tab:ReS}.
The background brightness of Fig.~\ref{fig:reg_beta} actually
indicates the value of $\ReS[z]$ (in arbitrary unit).
We only list the values at $\sigma_{0,1,2,\bar 0}$ 
because the critical points which interchange with each other
by the reflection about the real and imaginary axes
have the same $\ReS[\sigma_i]$ 
at $\hat \mu=0$ for real $\beta$.

The value $\ReS[\sigma_{\bar 0}]$ of the doubler solution is larger than 
$\ReS[\sigma_0]$ by $2\beta L=24$ for $n_-=0$
and $2 \beta (L-2)=12$ for $n_-=1$.
This difference comes from 
the bosonic part $\beta (1-\cos z)$ of the action.
On the other hand, 
the action $\ReS [\sigma_0]$
at $\sigma_0$ in $n_-=1$ sector is larger than
that in $n_-=0$ sector by a factor of order $2\beta=6$ 
because the former point contains one doubler component
$z_n=\pi$.
One may notice that 
$\ReS[\sigma_1]$ in $n_-=1$ sector takes a smaller value 
than $\ReS[\sigma_0]$, indicating the larger weight for it.
But $\K{\sigma_1}$ has no intersection with $\U$ at $\mu=0$,
and the thimble $\J{\sigma_1}$ is not a member of 
the integration cycles for $Z$.

It is intriguing to check this behavior with changing the lattice size
$L$ towards the continuum limit. 
By increasing $L$ and $\beta$ with keeping $\beta/L$ and $L ma$ fixed, 
there appear more zero points and accordingly the critical points
aligned in two rows. We compute $\ReS[\sigma_i]$ and list 
the results for $L=8$ and 16 
in the lower part of Table \ref{tab:ReS}. 
We observe that the contributions
from the $n_-=1$ sector to $Z$ 
are more suppressed by the factor
$e^{-2\beta}$ for the larger $L$ and $\beta$.
Within the $n_-=0$ sector, 
we can estimate the difference between $\ReS[\sigma_1]$ 
and $\ReS [\sigma_2 ]$ as $4\pi^2 \beta / L$ for larger $L$, basing on 
the bosonic part $L\beta(1-\cos z)$ 
and expanding it with approximation
$\sigma_k \sim z_{{\rm zero},k} \equiv (2k-1)\pi / L-\imag m$.
This gives us a factor $3\pi^2\sim 30$, 
which is consistent with the numerical result
of Table \ref{tab:ReS}.
For the smaller $\beta/L$ we have the smaller gap between
$\ReS [\sigma_1 ]$ and $\ReS \sigma_2 ]$.
The difference between $\ReS [\sigma_0 ]$ and $\ReS [\sigma_1 ]$ is
more sensitive to the choice of parameters $\beta/L$ and $Lma$.

In summary, 
we have clarified the thimble structure of the
Thirring model in this section.
The determinant zeros form submanifolds with complex dimension $L-1$,
and their sections in the subspace assigned with $n_-$
appear as isolated zero points.
The critical points of the model are classified with
$n_-$, and each of them pairs up with a zero point
in the subspace assigned with $n_-$
(except for the Gaussian critical point $\sigma_0$
and its doubler counterpart $\sigma_{\bar 0}$).
Thus all the thimbles are identified in the 
(0+1) dimensional Thirring model.
Towards the continuum limit ($\beta \to \infty$),
$\ReS[\sigma]$ with nonzero $n_-$,
which contains $n_-$ ``doubler'' components,
acquire the large values of order $2 n_- \beta$ compared 
to $\ReS[\sigma_0]$ in the $n_-=0$ subspace. 
This implies that the relative weights of their contributions to $Z$ 
are strongly suppressed toward the continuum limit,
even when they join the set of the 
integration cycles as $\mu$ increases.

\section{Stokes phenomenon and structure change at finite $\mu$}
\label{sec:Stokes}

%

In this section, with increasing $\mu$, 
we study the change of the intersection numbers and
thimbles which contribute to the partition function $Z$
from the viewpoint of the Stokes phenomenon and jumps.
We restrict our discussion in the uniform configuration space $n_-=0$.

\begin{figure}[tb]
\noindent
\includegraphics[width=35mm,bb=0 20 259 257]
{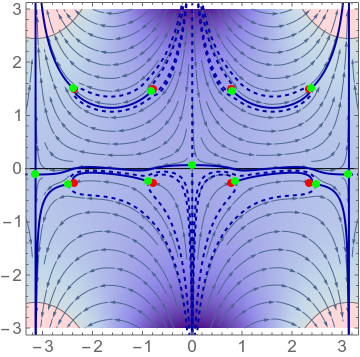}
\hfil
\includegraphics[width=35mm,bb=0 20 259 257]
{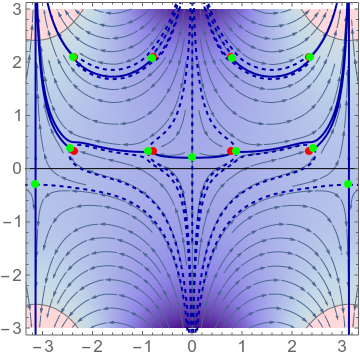}
\hfil
\includegraphics[width=35mm,bb=0 20 259 257]
{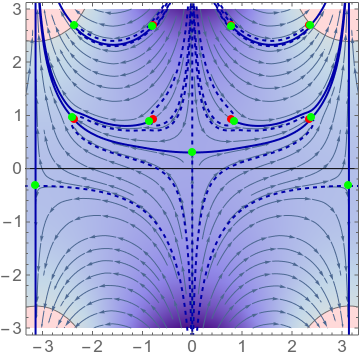}
\begin{center}
\hfil (a) $\hat \mu=0.6$
\hspace{3cm} (b) $\hat\mu=1.2$
\hspace{3cm} (c) $\hat\mu=1.8$
\hfil
{~}  
\end{center}
\vspace*{-5mm}
\caption{
Downward gradient flows, critical points (green) and zero points (red) 
in complex $z$ plane of the Thirring model with $ma=1$, $\beta=3$ and $L=4$
for (a) $\mu=0.6$,  (b) 1.2, and (c) 1.8.
The downward (upward) cycles, $\J{\sigma}$  ($\K{\sigma}$), are
depicted with solid (dashed) lines.
\label{fig:conf_zero}
}
\end{figure}

Figure~\ref{fig:conf_zero} shows 
the downward gradient flows of the model with $L=4, \beta=3$
and $ma=1$ for $\hat \mu= 0.6$, 1.2 and 1.8.
The zero points $z_{\rm zero}$'s and their associated critical points 
$\sigma$'s move upward as $\mu$ increases.
Then the critical points which align on the lower side,
cross the real axis at certain values of $\hat \mu \sim \hat m$ 
(see Eq.~(\ref{eq:zzero0})),
and accordingly the intersection numbers of 
${\cal K}_\sigma$'s with $\U$ change on the way. 
Now one encounters the situation where
certain thimbles join and/or leave the set of integration cycles for 
the partition function $Z$.
For a large enough $\hat \mu$,
as can be inferred from Fig.~\ref{fig:conf_zero}~(c),
the single thimble ${\cal J}_{\sigma_0}$ comes to connect the two safe remote
points $z=\pm \pi +i \infty$, 
to become an equivalent cycle to the
original U-shaped cycle: 
$\U \sim {\cal J}_{\sigma_0}$.

The downward and upward cycles $\J{\sigma}$ and $\K{\sigma}$ of
a critical point $\sigma$ generally extend to ``safe'' and ``unsafe''
remote regions, respectively,
without crossing other cycles $\J{\sigma'}$ and $\K{\sigma'}$ 
which have different values of ${\rm Im}S$. 
When multiple critical points share the same value of 
$\ImS$, the cycle associated to one of those critical points
may meet another critical point.
This is the so-called Stokes phenomenon.
Change of the intersection number is achieved only 
by a jump of one endpoint of a upward cycle $\K{\sigma}$ from (e.g.) 
$z = -\imag \infty$ to $z = \imag \infty$, and
in between the critical point $\sigma$ must undergo 
the Stokes phenomenon with another critical point $\sigma'$
in which $\K{\sigma}$ and $\J{\sigma'}$ just overlap.

As has been discussed in the previous section,
zeros of the fermion determinant become endpoints of the thimbles.
Because the determinant appears as $- N_f\log \det D$ 
in the action~$S$, the imaginary part 
$\ImS$ changes by $-2\pi N_f$ when we encircle a zero point 
counterclockwise from one side to the other side of a thimble 
which terminates at this zero.
However this difference is
not reflected in the gradient flow.
Therefore 
the necessary condition for the Stokes phenomenon to occur
between critical points $\sigma$ and $\sigma'$  is
\begin{align}
\ImS_{\sigma} = \ImS_{\sigma'} + 2 \pi k \qquad k \in {\mathbb Z}
\; .
\label{eq:ImS_ImS2}
\end{align}
Incidentally, the imaginary part $\ImS$ 
on the upward cycle (e.g.,) $\K{\sigma}$ may differ
by a multiple of $2\pi$ depending on which side of the thimble
$\J{\sigma}$ the cycle starts.
Moreover, 
since the value of $\ImS$ changes around a zero point,
two thimbles can meet at the zero point 
making an angle determined by the difference of their
$\ImS(\sigma_i)$. Thus,
one can read the relative phase of the two thimbles from their 
relative angle when they meet at the zero point.

\subsection{Stokes jumps with increasing $\mu$
\label{subsec:jump}
}

Let us study the Stokes phenomenon with increasing $\mu$ in more details.
We set $N_f=1$.
Because the configuration subspace for real $\beta$
is symmetric under reflection about the imaginary axis 
as seen in Fig.~\ref{fig:conf_zero},
we discuss the thimble structure on
the right-half plane hereafter. 
Even at finite chemical potential $\mu \ne 0$, 
this reflection symmetry $z \to - \bar z$
guarantees the realness of $Z$;
the thimbles which interchange under this transformation
give the contributions which are complex conjugate to
each other and whose sum becomes real\cite{Tanizaki:2015pua}.

\begin{figure}[tb]
\includegraphics[width=50mm]
{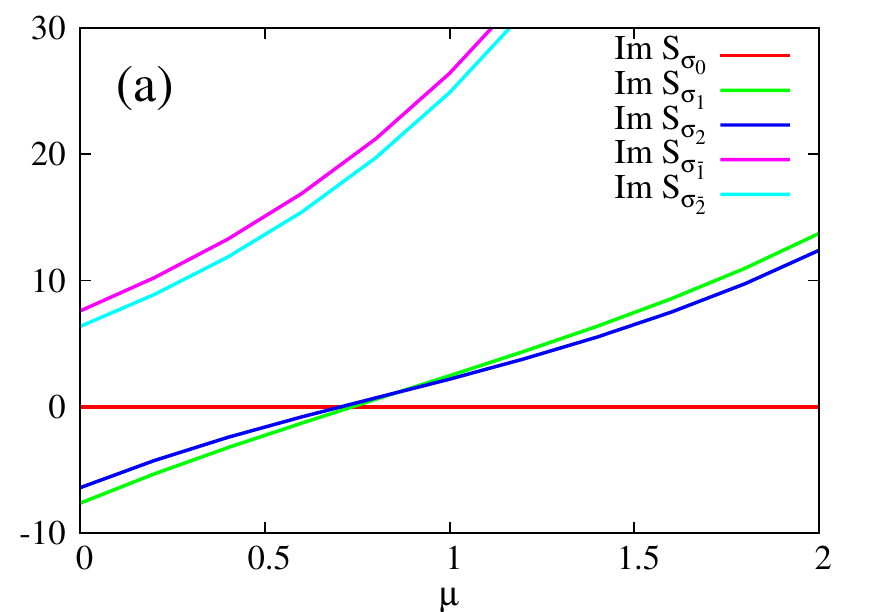}~\includegraphics
[width=50mm]
{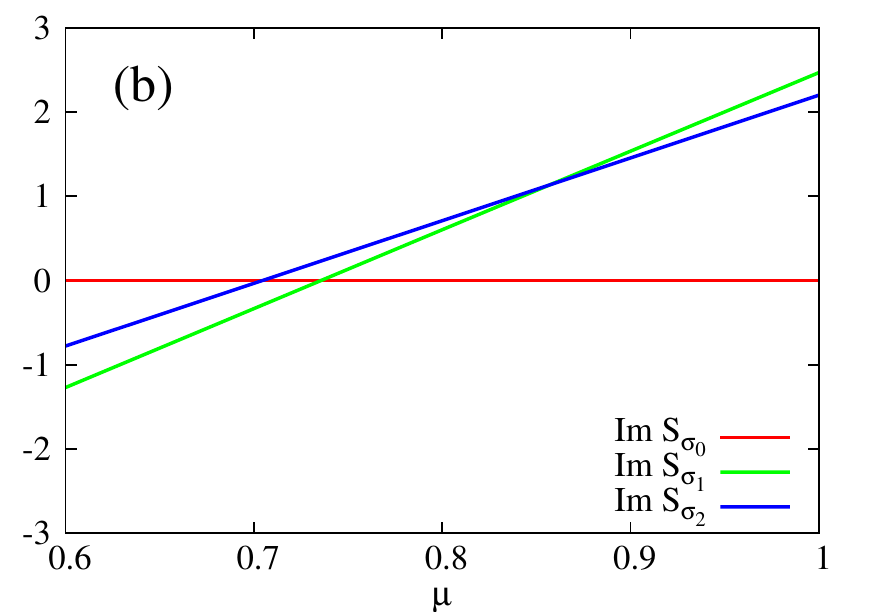}~\includegraphics
[width=50mm]
{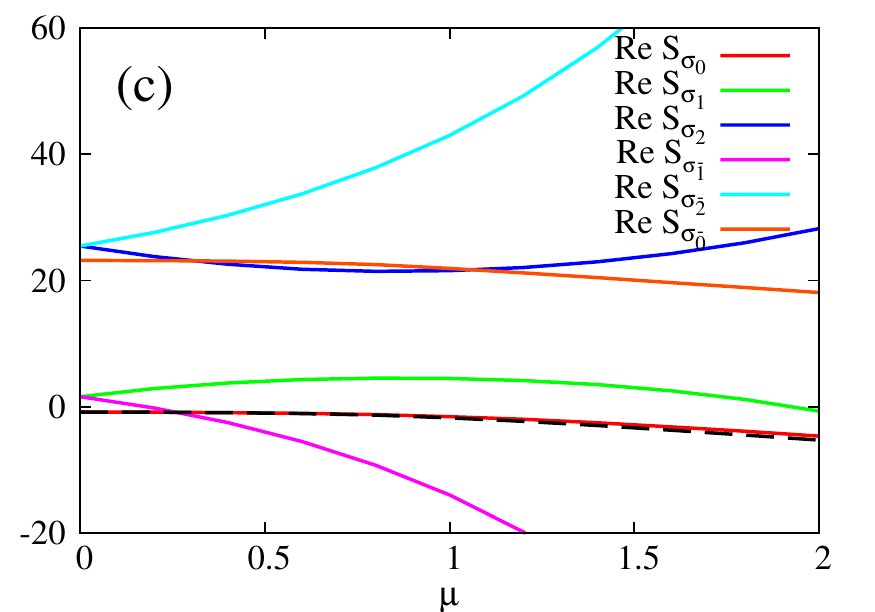}
\caption{
(a) $\ImS(\sigma_i)$ on the right half plane as a function of $\hat \mu$. 
(b) Enlarged plot of (a).
(c) $\ReS(\sigma_i)$. The dashed line indicates min.$_{x\in {\mathbb R}} \ReS(x)$.
Parameters are set to $L=4$, $\beta=3$ and $ma=1$.
}
\label{fig:ImS_m1_b3}
\end{figure}


In Fig.~\ref{fig:ImS_m1_b3}, we compare 
the values of the action at the critical points $\sigma_i$.
We first note that $\ImS=0$ at $\sigma_{0}$ and
$\sigma_{\bar{0}}$ independently of the chemical potential $\mu$.
Indeed, in Fig.~\ref{fig:conf_zero}~(a),
we see the Stokes phenomenon between $\sigma_{0}$ and
$\sigma_{\bar{0}}$, where the cycles 
$\J{\sigma_{0}}$ and $\K{\sigma_{\bar 0}}$ overlap,
and 
\begin{align}
\U \consim \J{\sigma_{0}} + \J{\sigma_{\bar{0}}}. 
\end{align}

At $\mu=0$ the critical points $\sigma_{\bar i}$ on the upper side 
have positive values of $\ImS(\sigma_{\bar i})$ and their associated
upward cycles $\K{\sigma_{\bar i}}$ extend to the unsafe region toward 
$z=+\imag \infty$.
With increasing $\mu$ the values of  $\ImS(\sigma_{\bar i})$
increase  monotonically and $\K{\sigma_{\bar i}}$ continue to
have no intersection with $\U$.
On the other hand, the critical points $\sigma_{i}$ on the lower side
move upward and the imaginary parts $\ImS(\sigma_{i})$ at these points
increase from negative to positive values with increasing $\mu$,
as seen in Fig.~\ref{fig:ImS_m1_b3}~(a).
In the enlarged plot in the panel (b), 
the lines of $\ImS_{\sigma_i}$ show three crossings 
at $\hat \mu=\hat \mu^*_1=0.7$, $\hat \mu^*_2=0.735$ and $\hat \mu^*_3=0.86$.
We now discuss the Stokes phenomenon and the change of the intersection
numbers at each $\hat \mu^*_i$.

\begin{figure}[tbhp]
  \begin{minipage}{0.48\hsize}
    \begin{center}
\includegraphics[width=45mm,bb=0 0 259 247]{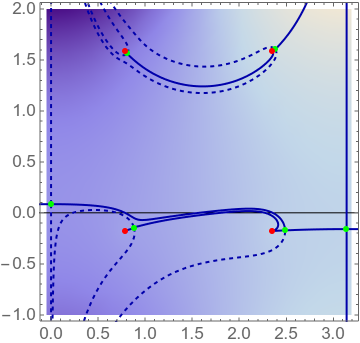}\\
          \hfil (a) $\hat \mu=\hat \mu^{*}_{1}$ \hfil 
       \end{center}
    \end{minipage}
\hfil
    \begin{minipage}{0.48\hsize}
      \begin{center}
\includegraphics[width=45mm,bb=0 0 259 247]{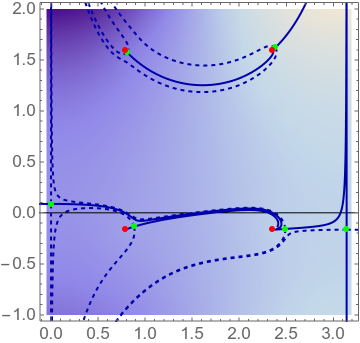}\\
          \hfil (b) $\hat \mu^{*}_{1}<\hat \mu<\hat \mu^{*}_{2}$ \hfil 
        \end{center}
      \end{minipage} 

\vspace{2cm}
      \begin{minipage}{0.48\hsize}
        \begin{center}
\includegraphics[width=45mm,bb=0 0 259 247]{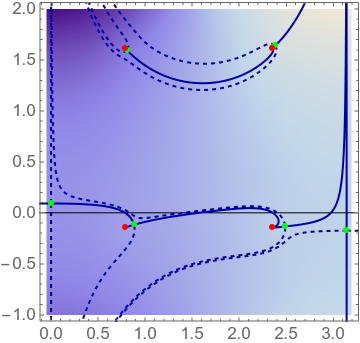}\\
          \hfil (c) $\hat \mu = \hat \mu^{*}_{2}$ \hfil
        \end{center}
      \end{minipage} 
\hfil
      \begin{minipage}{0.48\hsize}
        \begin{center}
\includegraphics[width=45mm,bb=0 0 259 247]{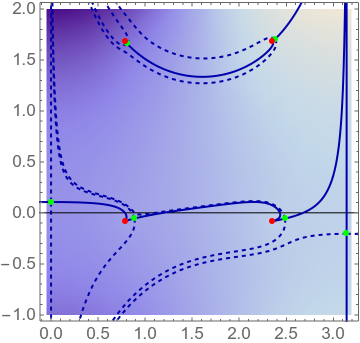}\\
          \hfil (d) $\hat \mu^{*}_{2}<\hat \mu<\hat \mu^{*}_{3}$ \hfil
        \end{center}
      \end{minipage} 

\vspace{2cm}
      \begin{minipage}{0.48\hsize}
        \begin{center}
\includegraphics[width=45mm,bb=0 0 259 247]{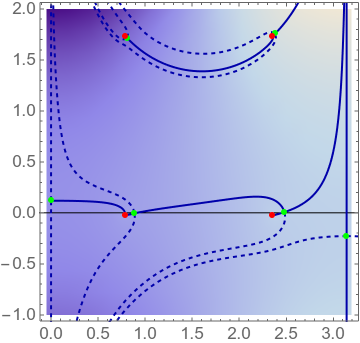}\\
          \hfil (e) $\hat \mu = \hat \mu^{*}_{3}$ \hfil 
        \end{center}
      \end{minipage} 
\hfil
      \begin{minipage}{0.48\hsize}
        \begin{center}
\includegraphics[width=45mm,bb=0 0 259 247]{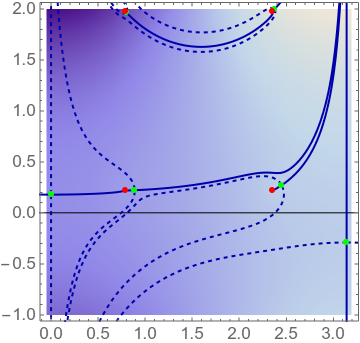}\\
\hfil    (f) $\hat \mu^{*}_{3}<\hat \mu< \hat \mu^{*}_{4}$  \hfil
        \end{center}
      \end{minipage} 
      \caption{Stokes phenomena at $\hat \mu=\hat \mu_i^*$ ($\hat \mu < \hat \mu_{4}^{*}$)}
  \label{fig:stokes}

\end{figure}

\begin{figure}[tbhp]
  \begin{minipage}{0.48\hsize}
    \begin{center}
      \includegraphics[width=45mm,bb=0 0 259 247]{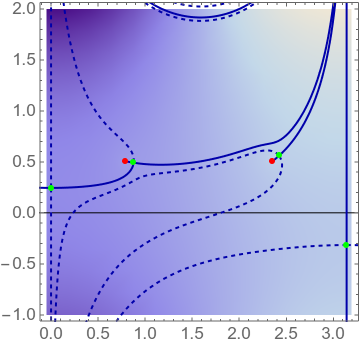}\\
\hfil  (g) $\hat \mu = \hat \mu^{*}_{4}$ \hfil 
    \end{center}
  \end{minipage} 
\hfil
  \begin{minipage}{0.48\hsize}
    \begin{center}
      \includegraphics[width=45mm,bb=0 0 259 247]{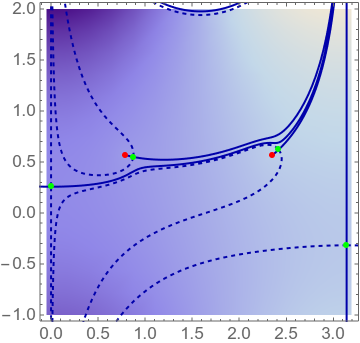}\\
\hfil (h) $\hat \mu^{*}_{4}<\hat \mu< \hat \mu^{*}_{5}$ \hfil 
    \end{center}
  \end{minipage} 

\vspace{2cm}
  \begin{minipage}{0.48\hsize}
    \begin{center}
      \includegraphics[width=45mm,bb=0 0 259 247]{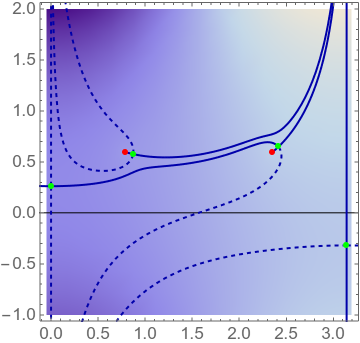}\\
\hfil (i) $\hat \mu=\hat \mu^{*}_{5}$ \hfil 
    \end{center}
  \end{minipage} 
\hfil
  \begin{minipage}{0.48\hsize}
    \begin{center}
      \includegraphics[width=45mm,bb=0 0 259 247]{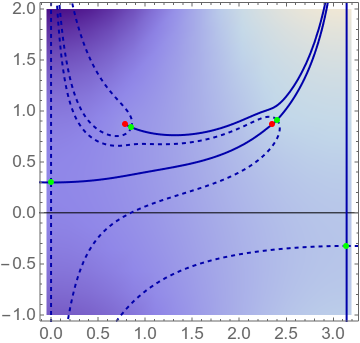}\\
\hfil (j) $\hat \mu^{*}_{5}<\hat \mu< \hat \mu^{*}_{6}$ \hfil  
    \end{center}
  \end{minipage}

\vspace{2cm}
  \begin{minipage}{0.48\hsize}
    \begin{center}
      \includegraphics[width=45mm,bb=0 0 259 247]{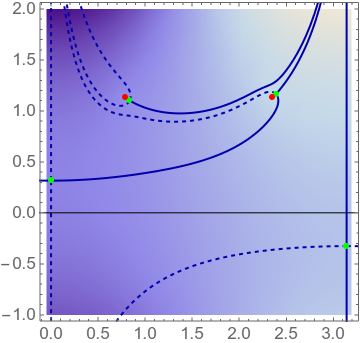}\\
\hfil (k) $\hat \mu=\hat \mu^{*}_{6}$  \hfil 
    \end{center}
  \end{minipage} 
\hfil
  \begin{minipage}{0.48\hsize}
    \begin{center}
      \includegraphics[width=45mm,bb=0 0 259 247]{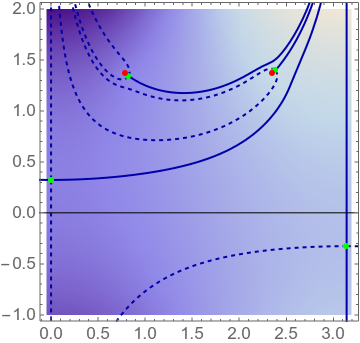}\\
\hfil (l) $\hat \mu^{*}_{6}<\hat \mu$  \hfil
    \end{center}
  \end{minipage}

  \caption{Stokes phenomena at $\hat \mu=\hat \mu_i^*$ ($\hat \mu_{4}^{*}  \le \hat \mu $)}
  \label{fig:stokes2}
\end{figure}

In Fig.~\ref{fig:stokes} we show typical thimble structures at several
values of $\hat \mu$.
At $\hat \mu < \hat \mu^*_1$, 
the cycles $\K{\sigma_{1,2}}$ starting from $\sigma_{1,2}$ 
extend to the lower unsafe  region toward $z= -\imag \infty$,
while 
$\K{\sigma_{\bar 1, \bar 2}}$ extend to the upper unsafe region toward
$z= + \imag \infty$.
None of them has nonzero intersection with $\U$,
and $\U \consim \J{\sigma_0}+\J{\sigma_{\bar 0}}$ as was discussed previously.
At $\hat \mu = \hat \mu^*_1$, 
$\ImS_{\sigma_0}=\ImS_{\sigma_2}$ is achieved,
and the two cycles $\J{\sigma_0}$ and $\K{\sigma_2}$ overlap.
Across $\hat \mu^{*}_1$,
one end of the upward cycle $\K{\sigma_2}$ jumps from $- \imag \infty$
to $+ \imag \infty$,
to give the intersection number $n_2=1$ with $\U$ (see panel (b)).
(And one end of the cycle $\J{\sigma_0}$ jumps from $\sigma_{\bar 0}$
to $z_{\rm zero, 2}$.)

At the same value of $\hat \mu=\hat \mu^*_1$,
the point $\sigma_2$ shows the Stokes phenomenon with another
critical point $\sigma_{\bar 0}$ 
because 
$\ImS_{\sigma_0} =\ImS_{\sigma_{\bar 0}} =0$. (This coincidence could be
avoided by adding a small imaginary part to $\beta$, again.)
In this case, the two cycles,
$\J{\sigma_2}$ and $\K{\sigma_{\bar 0}}$ overlap, and 
one end of the cycle $\J{\sigma_2}$ jumps from $\pi-\imag \infty$ 
to $\pi+ \imag \infty$ across $\hat \mu=\hat \mu^*_1$.
Hence, we have the equivalence of the cycles%
\footnote{
We define the orientation of a thimble as the direction where ${\rm Re}z$
increases, and define it for thimble $\J{\sigma_{\bar 0}}$ on the left as the
direction of decreasing ${\rm Im}z$.}  
\begin{align}
\U \consim  \J{\sigma_{-2}} + \J{\sigma_{0}} + \J{\sigma_{2}} 
\qquad \text{for~} \hat \mu^*_1 < \hat \mu < \hat \mu_2^* \; . 
\end{align}

At $\hat \mu=\hat \mu^{*}_{2}$ (panel (c)), 
the Stokes phenomenon happens between the
critical points, $\sigma_0$ and $\sigma_1$.
The two cycles $\J{\sigma_0}$ and $\K{\sigma_1}$ overlap there.
When $\hat \mu$ passes $\hat \mu^*_2$,
one end of the cycle $\K{\sigma_1}$ jumps from
$- \imag \infty $ to $+ \imag \infty$ and 
one end of the cycle $\J{\sigma_0}$ from 
$z_{\rm zero,2}$ to $z_{\rm zero,1}$, and therefore the critical point
$\sigma_1$ now acquires the intersection number $n_{\sigma_1}=1$.
Hence, 
\begin{align}
\U \consim
 \J{\sigma_{-2}}+ \J{\sigma_{-1}} 
+ \J{\sigma_{0}} 
+ \J{\sigma_{1}}  + \J{\sigma_{2}} 
\qquad \text{for~} \hat \mu^*_2 < \hat \mu < \hat \mu_3^* \; . 
\end{align}

At $\hat \mu=\hat \mu^{*}_{3}$ (panel (e)), 
$\ImS$ of $\sigma_1$ and $\sigma_2$ coincide, 
which allows the Stokes phenomenon between them.
Across $\hat \mu = \hat \mu^{*}_{3}$
one end of the cycle $\K{\sigma_2}$ flips
down from $ + \imag \infty$ to $ - \imag \infty$,
while one end of the cycle $\J{\sigma_1}$ jumps
from $z_{\rm zero,2}$ to $\pi + \imag \infty$,
so that the intersection number $n_2$ changes from 1 to 0.
Thus, we have ($\hat \mu^*_4$ introduced below)
\begin{align}
\U \consim  \J{\sigma_{-1}} + \J{\sigma_{0}} + \J{\sigma_{1}} 
\qquad \text{for~} \hat \mu^*_3 < \hat \mu < \hat \mu_4^* \; . 
\end{align}

So far we have discussed only the cases where the Stokes phenomenon
occurs between the critical points having the same value of $\ImS$.
For $\hat \mu$ larger than $\hat \mu^{*}_{3}$ we need to take into account the
multivaluedness of the logarithm because the edge of $\J{\sigma_0}$ is
now going around the zero points.
The condition for the Stokes phenomenon to occur is the
equality of $\ImS$ modulo $2\pi$
between the two critical points as announced in Eq.~(\ref{eq:ImS_ImS2}).
For our model parameters,
there are three more critical values $\hat \mu^*_{4,5,6}$. 
At $\hat \mu = \hat \mu^*_4$ (Fig.~\ref{fig:stokes2}~(g)),
the condition 
$\ImS_{\sigma_{0}} +2\pi =\ImS_{\sigma_{1}}$ is fulfilled,
and for $ \hat \mu^*_4 < \hat \mu < \hat \mu^*_5$ (Fig.~\ref{fig:stokes2}~(h))
the equivalent integration cycle becomes
\begin{align}
&\U \consim \J{\sigma_{0}} 
\qquad \text{for~} \hat \mu^*_4 < \hat \mu < \hat \mu_5^* \; . 
\end{align}
At $\hat \mu=\hat \mu^*_5$ (Fig.~\ref{fig:stokes2}~(i))
the condition 
$\ImS_{\sigma_{0}} + 2\pi = \ImS_{\sigma_{2}}$ is fulfilled,
and 
the equivalent integration cycle changes to
\begin{align}
&\U \consim   \J{\sigma_{-2}} + \J{\sigma_{0}} + \J{\sigma_{2}} 
\qquad \text{for~} \hat \mu^*_5 < \hat \mu < \hat \mu_6^* \; . 
\end{align}
At $\hat \mu = \hat \mu^{*}_{6}$ (Fig.~\ref{fig:stokes2}~(k)),
the condition
$\ImS_{\sigma_{0}} + 4\pi = \ImS_{\sigma_{2}}$ holds
and the equivalent integration cycle  now consists of a single thimble
\begin{align}
&\U \consim \J{\sigma_{0}}
\qquad \text{for~} \hat \mu^*_6 < \hat \mu \; . 
\end{align}

\begin{figure}[tbp]
\begin{center}
\includegraphics[clip, width=90mm]{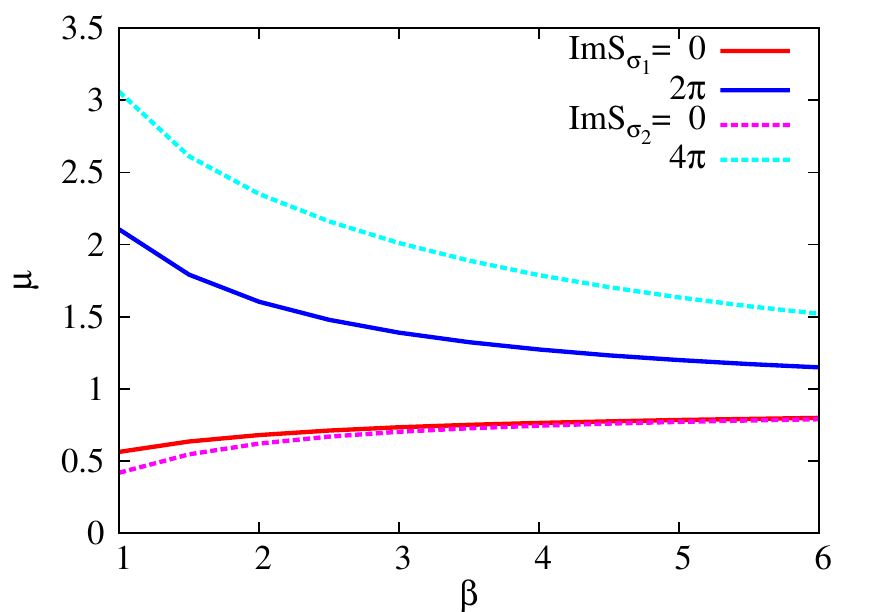}
\caption{
Critical values of $\hat \mu^*_{1,6}$ at which $\ImS_{\sigma_{2}}=0,4\pi$, 
and $\hat \mu^*_{2,4}$ at which $\ImS_{\sigma_{1}}=0,2\pi$,
      as a function of the coupling $\beta$ ($L=4$ and $ma=1$).}
    \label{fig:beta_dep}
\end{center}
\end{figure}

\subsection{Multi-thimble contributions and weight factor
\label{subsec:multithimble}
}

We have seen 
how the original integration cycle $\U$ is decomposed equivalently into  
a set of thimbles with increasing $\hat \mu$.
The partition function $Z$ is correctly reproduced
only if we evaluate the contributions from all the thimbles in the
set, in principle. 
Especially in the crossover region of $\hat \mu$, 
multiple thimbles take part in the set of the integration cycles.

However, importance of their contributions depends on the weight
factor $\exp[-\ReS(\sigma)]$.
For example, the integration
cycle consists of ${\cal J}_{\sigma_{0}}$ and $\J{\sigma_{\bar 0}}$ 
for $0 \le \hat \mu < \hat \mu^*_1$. 
But the contribution from $\J{\sigma_{\bar 0}}$
is numerically negligible because $\ReS(\sigma_{\bar 0})$ is larger
than $\ReS(\sigma_{0})$ by a large amount $\sim 2 L \beta $ 
as seen in Fig.~\ref{fig:ImS_m1_b3}~(c)
(see also Table \ref{tab:ReS} for $\hat \mu=0$ value).

For $\hat \mu^*_1 < \hat \mu < \hat \mu^*_4$
and $\hat \mu^*_5 < \hat \mu < \hat \mu^*_6$,
the thimbles $\J{\sigma_{\pm 1}}$ and/or $\J{\sigma_{\pm 2}}$ 
are in the set of the integration cycles 
in addition to $\J{\sigma_0}$.
According to the weight factor $\exp(-\ReS(\sigma))$ 
in Fig.~\ref{fig:ImS_m1_b3}~(c), 
the thimble $\J{\sigma_0}$ will give the largest contribution
and $\J{\sigma_{\pm 1}}$ will contribute as the second largest.
The contributions from $\J{\sigma_{\pm 2}}$ will be strongly suppressed.
This behavior is mainly controlled by the bosonic part 
$L\beta (1-\cos z)$ of the action.
(The thimble $\J{\sigma_{\bar 1}}$ is not
a member of the integration cycle, although
$\ReS(\sigma_{\bar 1})$ becomes smallest as $\hat \mu$ increases.)

In Fig.~\ref{fig:beta_dep} we plot 
the $\beta$ dependence of the critical chemical potential 
$\hat \mu^*_i$ for $L=4$ and $ma=1$.
Outside of the interval $\hat \mu^{*}_{1}<\hat \mu < \hat\mu^{*}_{6}$
the single thimble $\J{\sigma_0}$ becomes (almost) equivalent to the
original integration cycle $\U$, 
but within this interval multiple thimbles need to be considered.
Especially, the second-dominant thimbles $\J{\sigma_{\pm 1}}$ 
contribute in the interval $\hat \mu^{*}_{2}< \hat \mu < \hat \mu^{*}_{4}$.
We notice that the crossover region $\hat \mu \sim \hat m$
is indeed covered by this interval $\hat \mu^{*}_{2}<\hat \mu <
\hat\mu^{*}_{4}$, which indicates that the multi-thimble
contribution is requited to  reproduce the crossover behavior correctly.
The interval becomes wider (narrower) for smaller (larger) $\beta$.
From this $\beta$-dependence there may be a
possibility that the approximate evaluation of $Z$
with the single thimble $\J{\sigma_0}$ becomes better
for larger $\beta$.
Note that for larger $\beta$ 
the difference in the relative weights 
among the critical points also becomes more significant 
and the thimbles whose critical point locates away 
from $\sigma_0$ in the real axis direction is expected  to 
less contribute to $Z$.

In summary, for $L=4$ case,
we have clarified the change of the Lefschetz thimble structure
and the set of the thimbles contributing to $Z$
as $\hat \mu$ increases.
At small and large chemical potentials outside of the interval 
$\hat \mu^{*}_{2}<\hat \mu < \hat\mu^{*}_{4}$, 
the evaluation of $Z$ with the single thimble $\J{\sigma_0}$ is
legitimate, provided that $\J{\sigma_{\pm 2}}$ contributions are
negligibly small. But in the crossover region 
$\J{\sigma_{\pm 1}}$  contributions must be taken into account
in addition to that of $\J{\sigma_0}$.
The approximate evaluation by taking only one thimble
$\J{\sigma_0}$ is  performed in numerical simulations for several
models so far\cite{Cristoforetti:2013wha,Fujii:2013sra,Mukherjee:2014hsa,DiRenzo:2015foa}.
Hence it would be worthwhile to examine the validity of the single
thimble approximation across the crossover region
with varying $\beta$.
Furthermore
it would be intriguing to study how the crossover behavior
is reproduced by the multi-thimble contributions with 
increasing the lattice size $L$ toward the continuum 
and/or low temperature limits.

\section{Multi-thimble contributions in uniform-field model
\label{sec:ModelStudy}
}
In order to examine the single thimble approximation
and to investigate how the crossover behavior is reproduced
by contributions from multiple thimbles,
we study the Thirring model in the uniform-field subspace. 
The limitation to uniform-field configurations 
corresponds to the classical approximation with 
neglecting the quantum fluctuations. 
The partition function of this restricted model
is analytically evaluated to be
\begin{align}
 Z_{0} & = \int^{\pi}_{-\pi} \frac{d x}{2 \pi} \ 
\frac{1}{2^{L-1}} \, 
\Big [ \cosh (L(\hat \mu + \imag x)) + \cosh L\hat m \Big ]
\,  \e^{-L \beta(1-\cos x)}
\notag \\
& = \frac{e^{-\beta L}}{2^{L-1}}  \,
\big [ I_L(\beta L) \cosh L\hat \mu +  I_0(\beta L)  \cosh L \hat m
  \big ]
\, ,
\label{eq:Z0} 
\end{align}
and the fermion number density and chiral condensate are obtained
by
\begin{align}
\left < n \right >_0 &= 
\frac{1}{La}\frac{\partial}{\partial \mu} \log Z_{0},
\qquad
\left < \bar \chi \chi \right >_0 = 
\frac{1}{La}\frac{\partial}{\partial m} \log Z_{0}.
\end{align}
Interestingly, in the $T=0$ limit this classical model shows 
a first order transition at the same value 
of $|\mu_c|=m^2+g^2$ as the original model.

\subsection{Single-thimble approximation}

\begin{figure}[tbp]
\includegraphics[clip, width=0.48\textwidth]
{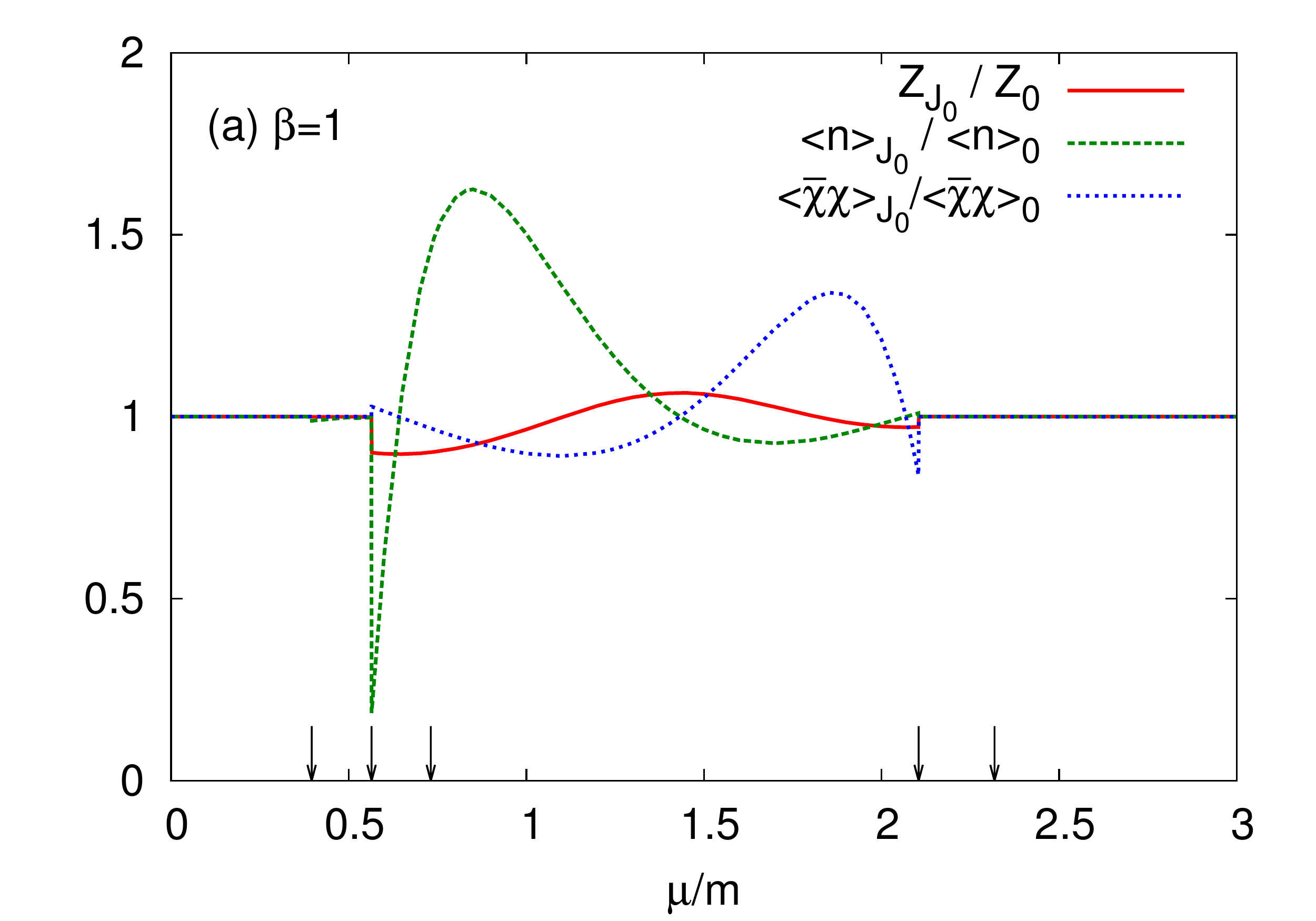}
\hfil  
\includegraphics[clip, width=0.48\textwidth]
{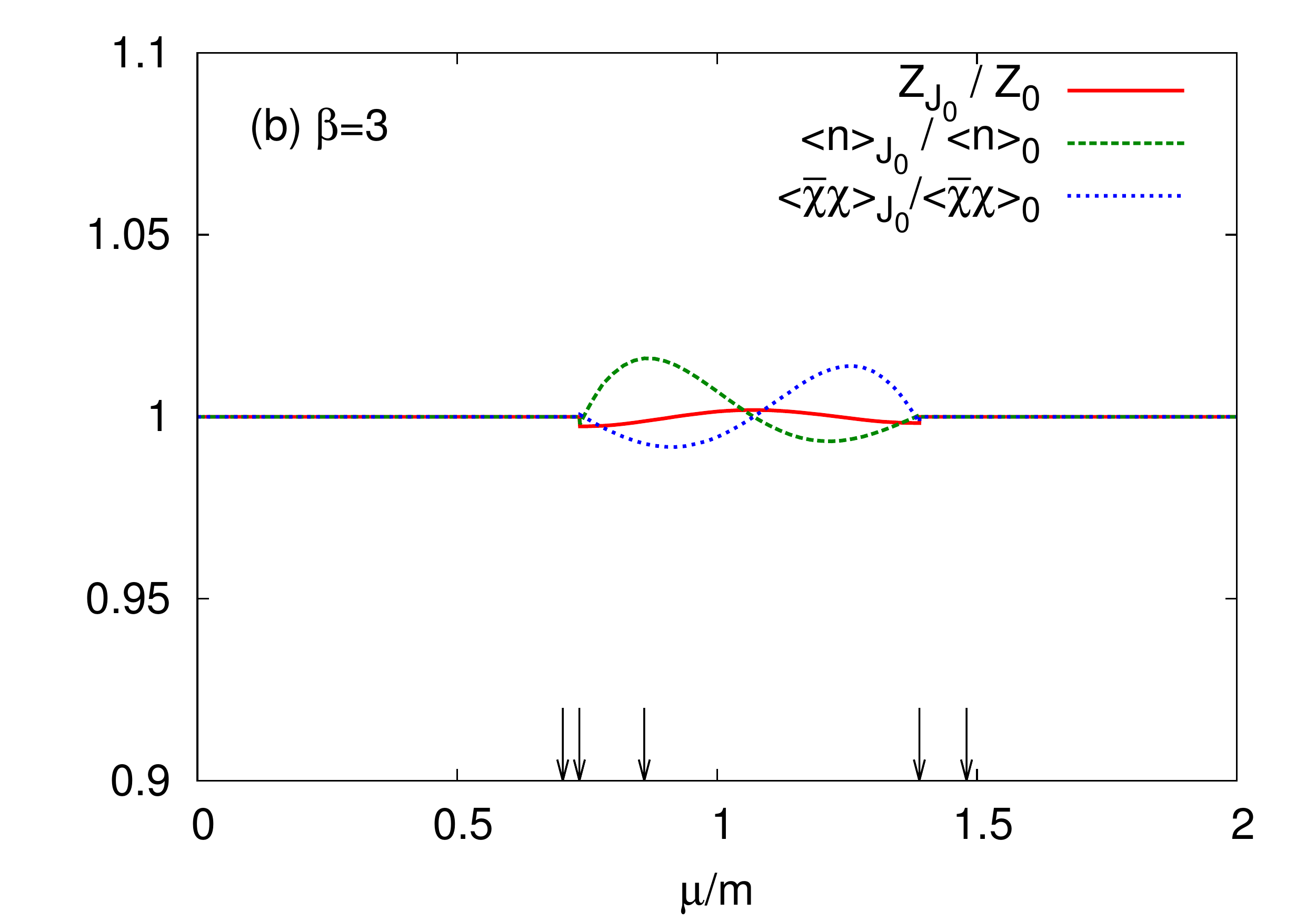}
\caption{
$Z_0$ (solid), $\left <n\right >_0$ (dashed)  and 
$\left <\bar \chi \chi \right >_0$ (dotted) evaluated on the single 
thimble ${\cal J}_{\sigma_{0}}$ normalized by the exact values 
of the uniform-field model 
for $\beta=1$ (a) and
3 (b) with $L=4$ and $ma=1$.
Arrows indicate the values of $\hat \mu^*_i$ ($i=1,\cdots,5$).}
\label{fig:Z0_Z}
\end{figure}

We compare the values evaluated on the single thimble $\J{\sigma_0}$
to the exact ones by taking their ratios in Fig.~\ref{fig:Z0_Z}.
We show the results with $L=4$ and $ma=1$ for $\beta=1$ (left) and 3 (right).
The critical values of the chemical potential
 $\hat \mu^*_{i=1, ..., 6}$ for the Stokes phenomenon
are found to be 
\{0.40, 0.56, 0.73, 2.10, 2.31, 3.0\} for $\beta=1$,
and 
\{0.70, 0.735, 0.86, 1.39, 1.48, 2.01\} for $\beta=3$.
We see that the single thimble integration gives us practically
the exact results
outside the region of $\hat \mu^*_2 < \hat \mu<\hat \mu^*_4$ in both cases.
This is because,
compared to the thimbles $\J{\sigma_0}$ and $\J{\sigma_{\pm 1}}$, 
the thimbles $\J{\sigma_{\pm 2}}$ and $\J{\sigma_{\bar 0}}$
have so small weight factor $\exp(-\ReS)$ that 
their participation in the integration cycle are numerically negligible.

On the other hand, 
the results deviate from unity in the range of $\hat \mu^*_2 < \hat \mu<\hat \mu^*_4$,
indicating that
the contributions from $\J{\sigma_{\pm 1}}$ need to be included 
to reproduce the original integral quantitatively. 
The much smaller deviation for $\beta=3$ case can
be understood if one recalls the rough estimate
for the weight factor
$\exp(-\ReS(\sigma_{\pm 1})) \sim \exp(-\beta \pi^2/(2L))$
as discussed in subsec.~\ref{subsec:thimble}.
Furthermore, we notice that the missing $\J{\sigma_{\pm 1}}$
contribution  to $Z$ changes the sign from positive to negative, 
and back to positive again, as $\hat \mu$ increases.
This is the reflection of the fact that
$\ImS(\sigma_1)$ increases from 0 at $\hat \mu=\hat \mu^*_2$
to $2\pi$ at $\hat \mu=\hat \mu^*_4$.
Because $\ImS(\sigma_0)=0$ for any $\hat \mu$,
the two thimbles $\J{\sigma_1}$ and $\J{\sigma_0}$ contribute additively
just above $\hat \mu=\hat \mu^*_2$.
But when $\ImS(\sigma_1)=\pi$,
they contribute with opposite signs.
At this point they are connected at $z=z_{\rm zero,1}$ 
with an angle $\pi$ between their edges as seen in
Fig.~\ref{fig:stokes} (f).
The $\J{\sigma_{\pm 1}}$ contributions return to be positive
as $\hat \mu$ approaches the critical value $\hat \mu^*_4$ for the
Stokes phenomenon.  
Regarding $\left < n \right >$ and $\left < \bar \chi\chi \right >$,
their integrands have non-constant imaginary parts on 
$\J{\sigma_{\pm 1}}$, and the contributions of $\J{\sigma_{\pm 1}}$ to
these densities alternate in different ways 
in the interval  $\hat \mu^*_2 < \hat \mu<\hat \mu^*_4$.

\subsection{Toward continuum limit}

\begin{figure}[tbp]
\includegraphics[clip, width=0.48\textwidth]
{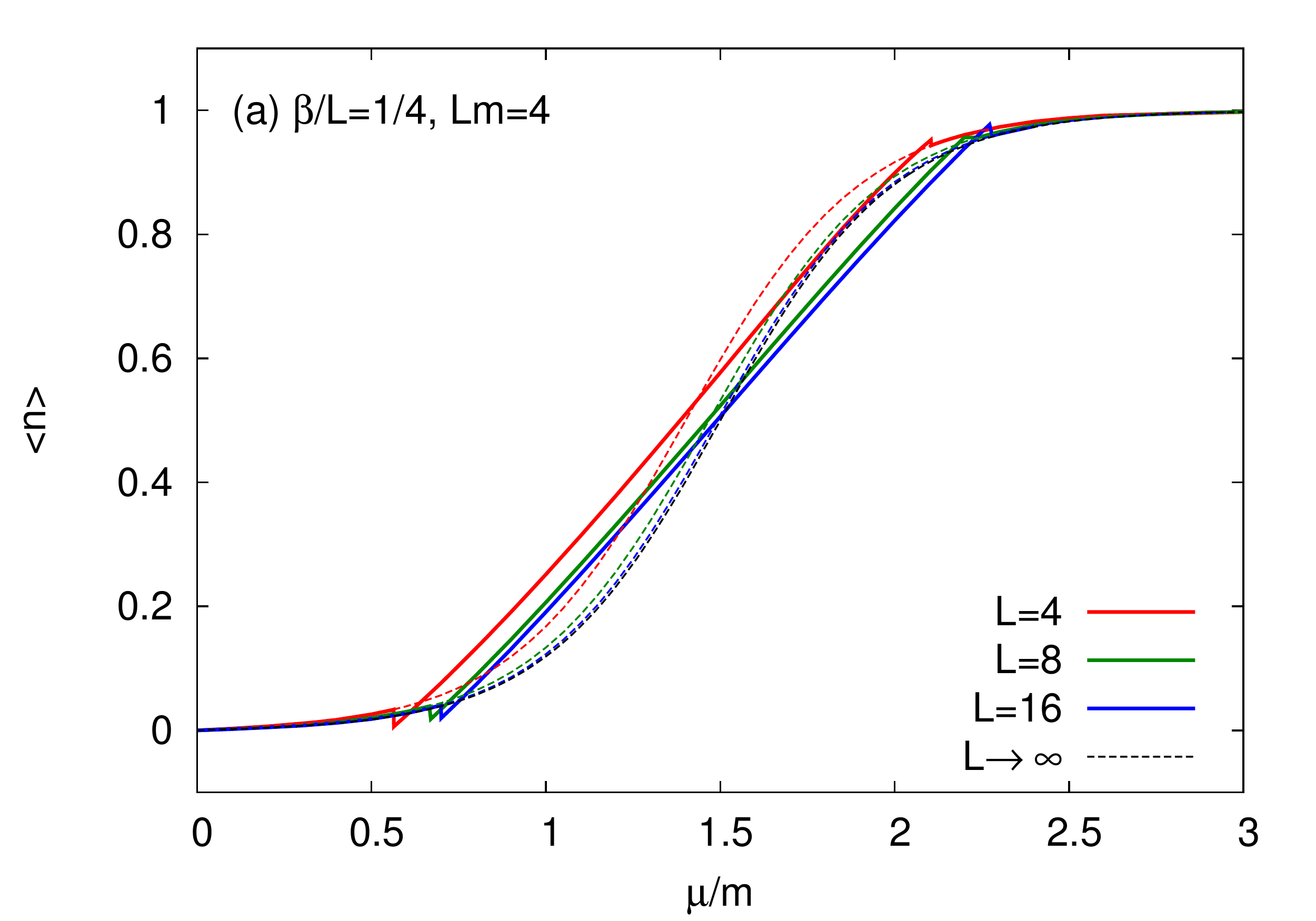}
\hfil  
\includegraphics[clip, width=0.48\textwidth]
{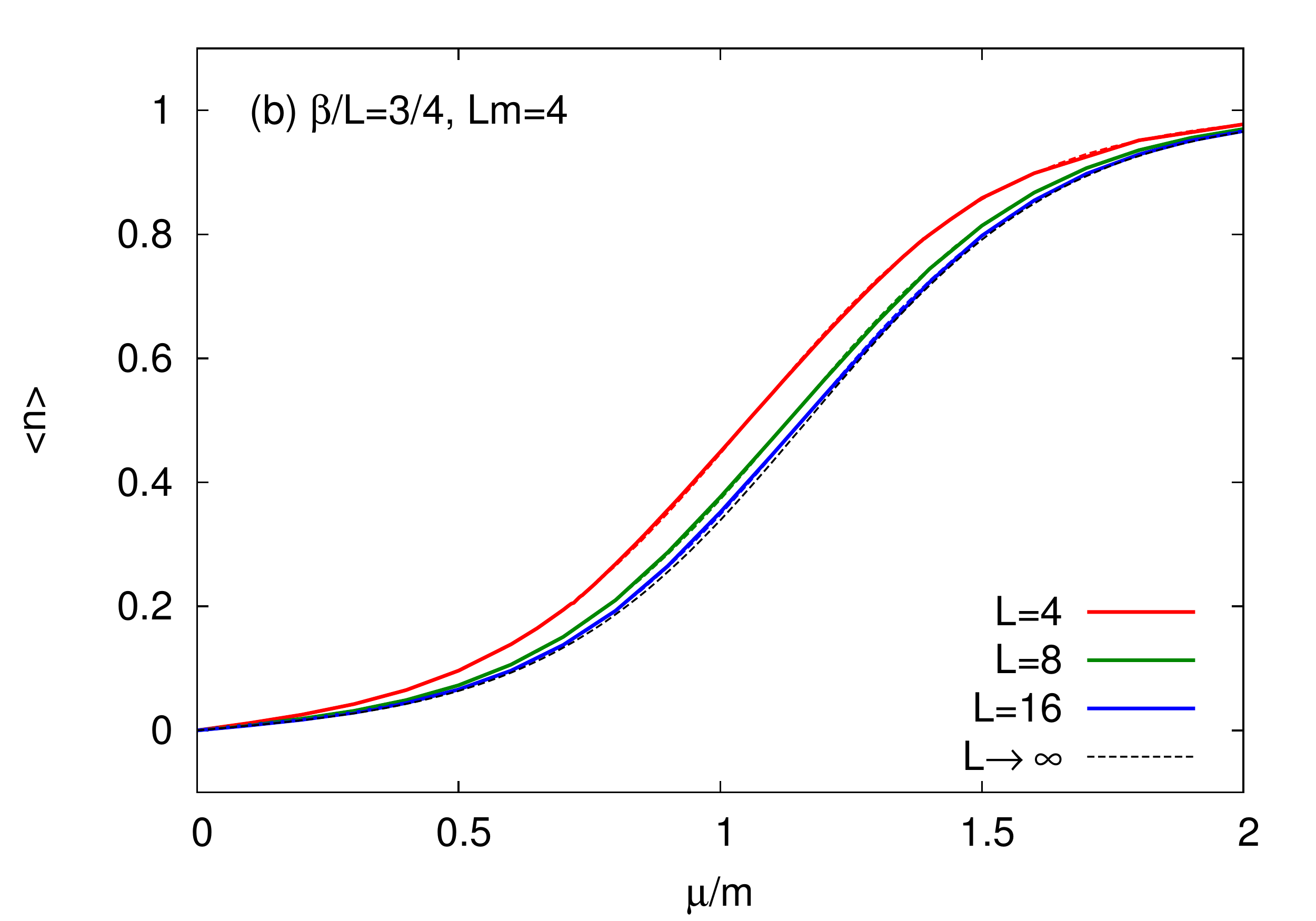}
\begin{center}
\caption{(a) Fermion number density as a function of $\hat \mu$,
evaluated on the single thimble $J_{\sigma_0}$ 
for $L=4$ (red), 8 (green), 16 (blue) 
with fixed $(\beta/L,Lm)=(1/4, 4)$.
(b) The same as (a) but with $(\beta/L,Lm)=(3/4, 4)$.
The uniform-field exact ones are shown in dashed lines for
comparison.
\label{fig:n-continuum}}
\end{center}
\end{figure}

In Fig.~\ref{fig:n-continuum} we examine the behavior of 
the fermion number density $\left < n \right >_{\J{0}}$ 
evaluated only on the single thimble $\J{\sigma_0}$
as a function of $\mu/m$
for $L=4,8,16$ toward the continuum limit.
The parameters are set to (a) $(\beta/L,Lm)=(1/4, 4)$
and (b)~$(\beta/L,Lm)=(3/4, 4)$.
In Fig.~\ref{fig:n-continuum}~(a), 
some discrepancy from the exact value (dashed line) 
is seen between 
$\hat \mu^*_2 < \hat \mu < \hat \mu^*_4$ for $L=4$, 
where the thimbles $\J{\sigma_{\pm 1}}$ have
the nonzero intersection number and need
to be included in the integration.
This behavior persists when 
we increase the lattice size to $L=8, 16$ 
toward the continuum limit
(thin black dashed curve).
The critical values $\mu^*_i/m$ for the Stokes phenomenon with
the thimbles $\J{\sigma_{\pm 1}}$ only slightly shift to larger $\hat \mu$
toward the continuum limit.
In Fig.~\ref{fig:n-continuum}~(b),
The discrepancy from the exact values is practically invisible
and again the results are relatively insensitive to the
size of the lattice with our parameters.
This implies that at finite temperatures Monte Carlo simulations on a
single thimble may work well for a certain parameters.

\subsection{Toward low temperature limit}

Next we change $L$ as 4, 8, and 16 with fixed $\beta=1$ and $ma=1$,
toward the zero temperature limit in Fig.~\ref{fig:nlowT}.
We find that the agreement between 
$\left < n \right >_{\J{0}}$ and 
$\left < n \right >_{0}$ is getting worse as $L$ increases.
Even in $\beta=3$ case (Fig.~\ref{fig:nlowT} (b))
we see a significant discrepancy from the exact result (dashed line)
for larger $L$.
As $L$ increases, the slope of the exact curve becomes
steeper in the crossover region and eventually 
converges to a step function, 
while the single thimble result 
$\left < n \right >_{\J{0}}$ behaves almost as a linear function
between two kink points.
The singular points indicate the Stokes jump
occurring there, through which the thimbles $\J{\sigma_{\pm 1}}$
join or leave the set of the integration cycles for the partition
function $Z$.  
 
\subsection{Multi-thimble contributions}

\begin{figure}[tbp]
\begin{center}
\includegraphics[width=70mm,bb=0 0 349 244]
{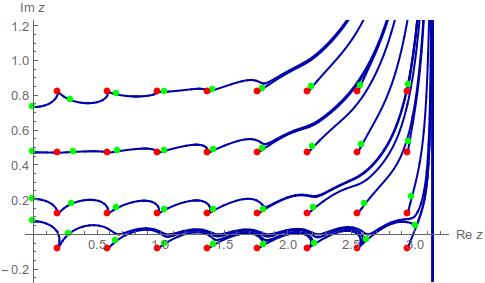}
\caption{Thimble structure on the right-half plane of $z$ 
with $L=16, \beta=1, ma=1$ 
for $\hat \mu=0.8, 1.0, 1.35, 1.7$ (from bottom to top).
The critical (zero) points are indicated with green (red) dots.
\label{fig:L16thimblestructure}
}
\end{center}
\end{figure}

We draw the thimble structure on the right-half plane
for $L=16$
at $\hat \mu=0.8, 1.0, 1.35, 1.7$ in the crossover region 
with $\beta=1, ma=1$ in Fig.~\ref{fig:L16thimblestructure}.
At $\hat \mu=0.8$ 
the three thimbles $\J{\sigma{0}}$ and $\J{\sigma{\pm 1}}$
have the nonzero intersectin numbers with the original integration
cycle, while 
at larger $\hat \mu$ the thimbles $\J{\sigma{0, \pm 1, \pm2, \pm 3, \pm 4}}$ 
(according to our numbering) intersect, 
and they need to be included as the integration cycles
to reproduce the partition function $Z_0$.

\begin{figure}[tbp]
\includegraphics[clip, width=0.48\textwidth]
{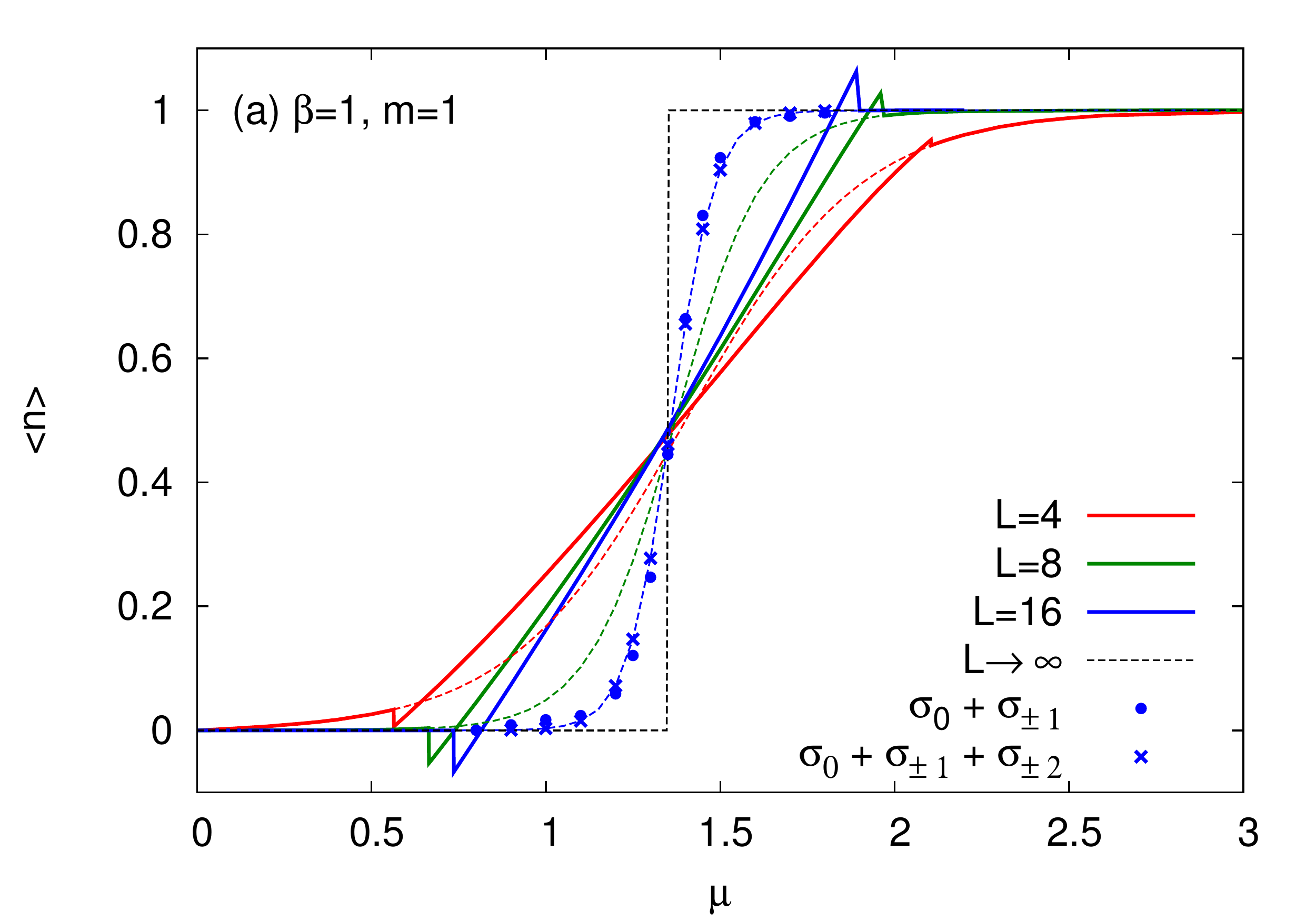}
\hfil  
\includegraphics[clip, width=0.48\textwidth]
{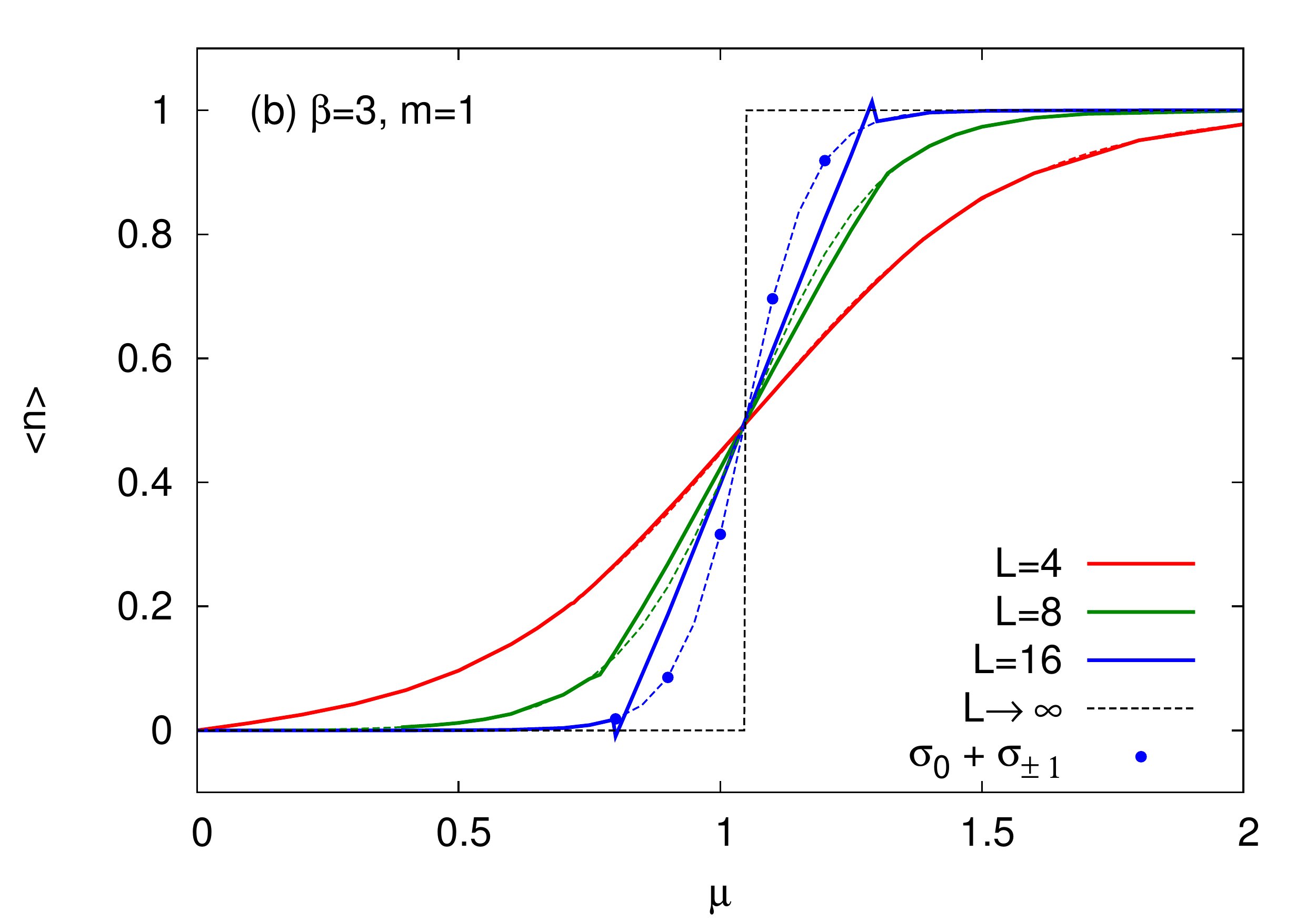}
\begin{center}
\caption{(a) Fermion number density as a function of $\hat \mu$,
evaluated on the single thimble $J_{\sigma_0}$ 
for $L=4$ (red), 8 (green), 16 (blue) 
with fixed $(\beta,m)=(1, 1)$.
(b) The same as (a) but with $(\beta,m)=(3, 1)$.
The uniform-field exact ones are shown in dashed lines.
For $L=16$, 
improved evaluations by including thimbles $\J{\sigma_{\pm1 (\pm 2)}}$
are shown with dots (crosses).
$L \to \infty$ limit is shown in a thin solid line.
\label{fig:nlowT}
}
\end{center}
\end{figure}

Based on this observation,
we extend the evaluation by including the contributions 
from $\J{\sigma_{\pm 1}}$ for $\beta=1, 3$
and those from $\J{\sigma_{\pm 2}}$ further for $\beta=1$,
as shown with dots and crosses in Fig.~\ref{fig:nlowT}.
Indeed, the agreement between the exact and
multi-thimble evaluations becomes systematically improved
by taking into acount the multi-thimble contributions.

In Table~\ref{tab:thimblecontrib} we listed the contributions to the
partition function $Z_0$  and 
the fermion density $\left < n \right >_0$ from each thimble
with $L=16$, $\beta=1$ and $ma=1$.
The thimbles $\J{\sigma_{\pm i}}$ give the contributions 
which are complex conjugate to each other so that their sum becomes
always real. 
Regarding partition function $Z_0$,
the thimble $\J{\sigma_{\pm 0}}$ gives the largest contribution,
but the thimbles $\J{\sigma_{\pm 1}}$ also provide a substantial
contribution in this crossover region.
Those from $\J{\sigma_{\pm i}}$ ($i \ge 2$) decrease
rather quickly as $i=2,3,4$ increases, which will be
very favorable for a systematic expansion.
But we notice that a cancellation occurs between the
$\J{\sigma_0}$ and $\J{\sigma_{\pm 1}}$ contributions at $\hat \mu=1.35$
owing to the negative sign of the $\J{\sigma_{\pm 1}}$ contributions.
For the fermion density $\left < n \right >_0$
the cancellation between the 
$\J{\sigma_0}$ and $\J{\sigma_{\pm 1}}$ contributions becomes more
delicate at $\hat \mu=0.8$ and $1.0$, while those come to contribute
additively at $\hat \mu=1.7$.
Insensitivity of the observables in small chemical region at low
temperatures, especially at zero temperature,
is sometimes called {\it Silver Blaze} phenomenon. 
We find here that when multiple thimbles contribute to the partition
function they show a delicate cancellation between them.

The alternating sign $\exp(-\ImS)$ of the thimbles at $\hat \mu=1.35$
manifest in Fig.~\ref{fig:L16thimblestructure} as the fact that 
the critical points and zero points are aligned and
the thimbles are connected at each zero point
with the angle about $\pi$.
In order to check this alternating pattern, 
we extend our calculation to $L=32$ as listed in the bottom row 
in Table~\ref{tab:thimblecontrib}.
We find that the thimble-by-thimble alternating sign and cancellation
become more striking not only for $Z_0$ but also for $\left < n \right >_0$.
In this case, we need to include the thimbles up to $\J{\sigma_{\pm 3}}$
to evaluate the observables with a few \% accuracy.
At larger $L$  more zero points appear near the imaginary
axis (Eq.~(\ref{eq:zzero0})),
and in between the critical points and associated thimbles 
are aligned at $\hat \mu$ in the crossover region. 
The weight factor from the bosonic part of the action does
not suppress these thimble contributions as far as 
${\rm Re} (L \beta (1-\cos \sigma_i))<1$.
Therefore we need to treat the neat cancellation in multiple thimble
contributions in order to reproduce the sharp rise of the fermion
density at low temperature (large $L$).
Implication of this observation to the feasibility of 
the numerical simulations with large lattice size is left for future study.

\begin{table}
\begin{center}
\footnotesize
\begin{tabular}{c|c|ccccc}
\hline
$\hat \mu$ & $Z_0$, $\left <n\right>_0$  & 0 & 1 & 2 & 3 & 4  \\
\hline
0.8&2.04 &   1.19     & (0.43, 0.04) & --- & --- & --- \\
   &1.3E-4&  -7.33E-3  & (3.73E-3, -7.351E-2)& --- & --- & --- \\
\hline
1.0 & 2.05 & 1.50 & (0.28, -0.42)&(-0.005, -0.021) & (-1E-4, -1E-4) & (-3E-7, -2E-7)\\
    & 3.2E-3 & 0.1186 & (-0.0508, -0.0774)&(-6.9E-3, 0.6E-3) & (-5E-5, 5E-5)  & (-9E-8, 2E-7)\\
\hline
1.35 &3.80  &9.09 & (-2.72, -0.39)&(0.07, 0.05) & (1E-3, -4E-4) & (-3E-7, -3E-7)\\
     &0.46  &1.17 &(-0.37, 0.23) &(0.016, -0.008) &(-9E-5, -4E-5)& (-1E-7, 8E-8)\\
\hline
1.7  &474.2 & 374.7  & (51.0, 80.7) &(-1.3, 0.9)  & (1E-3, -2E-3) & (-7E-7, -2E-7) \\
     &1.00  &0.67   &(0.16, 0.09)   &(-1E-4, 4E-3)  & (-4E-6, -5E-6)&(-7E-10, 2E-9)\\
\hline
\hline
1.35 &54.91  & 569.97 & (-298.63, -30.39) & (42.60, 13.20)&(-1.51,-1.27) & (5E-3, 2.8E-2)\\
     & 0.47 &  5.05 &  (-2.72, 0.84) & (0.45, -0.20) & (-0.025, 6.6E-3) &  (4E-4, 1E-4) \\
\hline
\end{tabular}
\caption{Contributions of thimbles on the right-half plane to $Z_0$
  (upper) and to $\left < n\right >_0$ (lower)
with $L=16$, $\beta=1$ and $ma=1$ for $\hat \mu=0.8, 1.0, 1.35, 1.7$. 
Thimbles on the left-half plane give the values complex conjugate to those in the list.
Below the double line, those values with $L=32$ are listed.
}
\label{tab:thimblecontrib}
\end{center}
\end{table}

\section{Summary and discussions
\label{sec:Summary}}

We have studied the Lefschetz thimble structure
of the (0+1) dimensional Thirring model at finite chemical potential,
which is formulated on the lattice of size $L$ 
with the staggered fermions and
a compact auxiliary vector field. This model suffers from the sign
problem by the complex fermion determinant.

The fermion determinant brings in two important features in the 
complexified field space: 
many isolated critical points of the gradient flow and submanifolds 
of the zero points with complex dimension $(L-1)$.
Those critical points accompany the Lefschetz thimbles and
the submanifolds of the zeros serve the ending points for the thimbles.
We have identified all the critical points of this model, 
and furthermore we have pointed out a one-to-one correspondence
between a critical point and a zero point within a projected
configuration subspace assigned with $n_-$.

We argued that the thimbles associated with the critical points in
$n_-=0$ subspace become more important toward the continuum limit
because the relative weights of the other critical points located
in $n_- \ne 0$ subspaces are suppressed by powers of $\e^{-2\beta}$.
The critical points with nonzero $n_-$ actually 
involve the doubler components and they are expected 
naturally to decouple from the system in the continuum limit.

Hence, restricting our analysis to the critical points in the $n_-=0$
subspace, 
we have shown how the thimble structure changes via the
Stokes jumps as the chemical potential $\mu$ increases.
We found that at small and large chemical potentials 
the single thimble $\J{\sigma_0}$ is sufficient as the integration
cycle to reproduce the partition function of the model.
However in the crossover region we must include multiple thimbles 
in the set of the integration cycles for the partition function $Z$.
Their relative weights depend on the lattice size $L$ and
the coupling strength $\beta$.

Taking the uniform-field model as a concrete example,
we have examined the importance of the multi-thimble contributions
and how the crossover behavior is generated by them.
The single-thimble approximation is justified 
for large $\beta/L \sim T/g^2$, even in the continuum limit.
But as we increase the lattice size $L$, 
i.e., lower the temperature $T$
with $\beta$ and $ma$ fixed, we have seen the breakdown of the
single-thimble approximation, which indicates the necessity of the
multi-thimble contributions. The sign of those contributions 
is alternating, which yields a neat cancellation
to reproduce the correct values of $Z$ and observables at large $L$. 
We notice that the contributions from the thimbles away from the origin
diminish rather quickly.
The Silver Blaze behavior
and the following
abrupt rise of the density $\left < n \right >$
with increasing $\mu$ 
are achieved by the interplay among the multi-thimble contributions
in the crossover region.

\begin{figure}[tbp]
\begin{center}
\includegraphics[width=90mm]{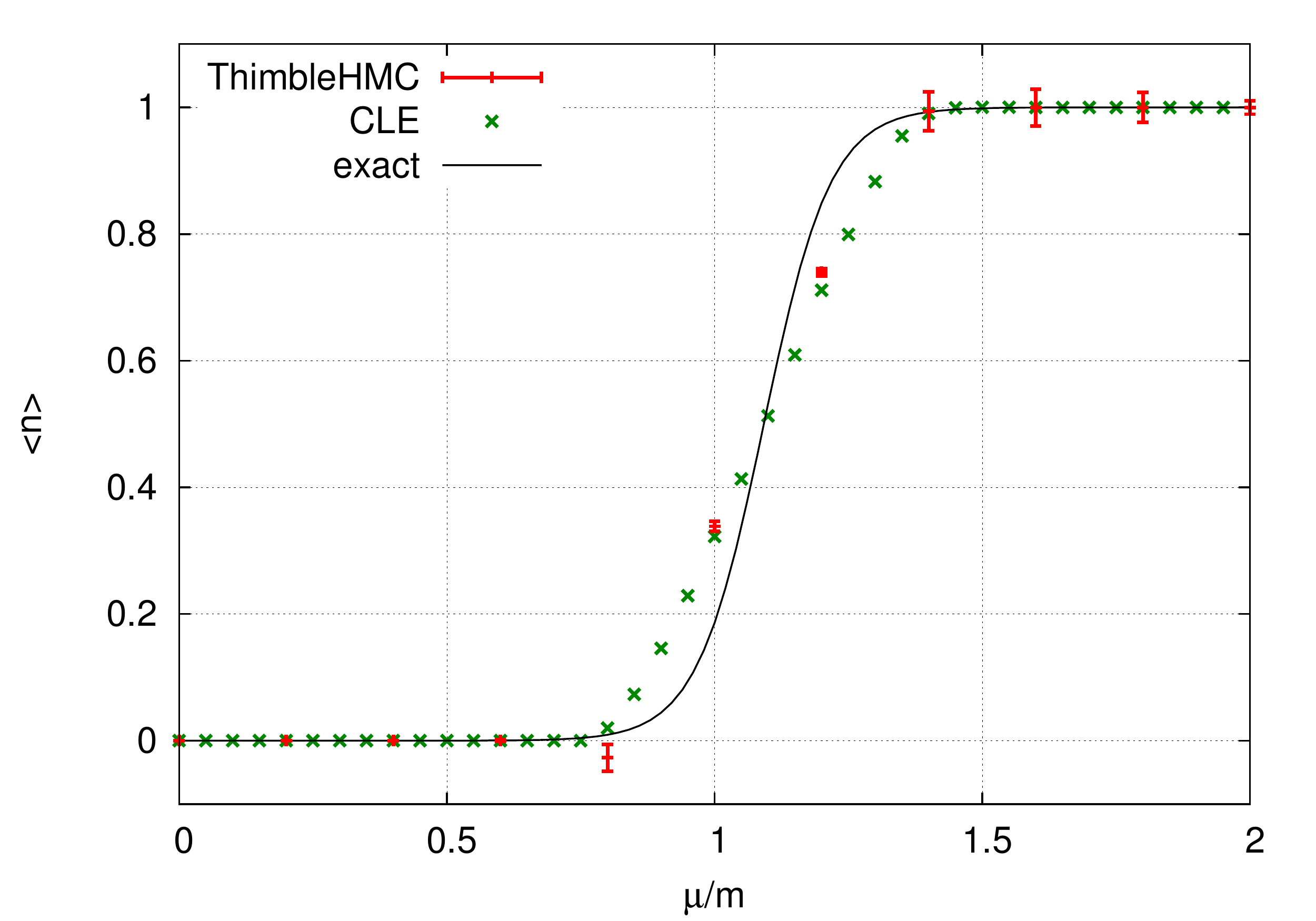}
\caption{Result of HMC simulation on the thimble $\J{\sigma_0}$
with $L=16$, $\beta=3$ and $ma=1$.
The curve indicates the exact value.
Result of complex Langevin simulation 
(step size $10^{-4}$, and $10^4$ samples taken every 100 steps)
is also shown for comparison.
\label{fig:ThimbleVsCLE}
}
\end{center}
\end{figure}

We have performed HMC simulations for the (0+1) dimensional Thirring
model with finite chemical potential 
on the single thimble $\J{\sigma_0}$ in Ref.~\cite{FKK2}.
We observed scaling behavior of the results to the continuum limit at finite
temperature and to the low-temperature limit.
The single thimble evaluation in the crossover region
is getting worse for smaller $\beta$ and/or
larger $L$, which is consistent with the results obtained in the
uniform-field model. 
We show one example of the simulation results for $L=16$, $\beta=3$
and $ma=1$ in Fig.~\ref{fig:ThimbleVsCLE}.

For comparison, 
we also tried the complex Langevin simulation as yet another approach
with complexification and as a possible way to include the
``multi-thimble'' contributions, which is shown in Fig.~\ref{fig:ThimbleVsCLE}.
We find that the Langevin result also deviates from the exact one
in the crossover region, but in a different manner. 
We observed that the sampling points in the Langevin simulation are
distributed around the thimbles $\J{\sigma_0}$
and $\J{\sigma_{\pm 1}}$.
The details of the Langevin simulation
will be reported elsewhere.

We have seen that an interplay among multi-thimble contributions are
necessary and important to describe the rapid crossover behavior of
the fermion system. 
However it is a difficult task to identify all the critical
points in generic models. Our analysis suggests that
the thimbles whose critical points locate in the uniform-field
subspace will give dominant contributions, while those with critical
points in non-uniform-field subspace will decouple by the suppressed weight
factor toward the continuum limit because they have doubler components.
Assuming that we can identify all the relevant thimbles to be integrated
over, we will face another challenge -- how to add up the
multi-thimble contributions in the Monte Carlo simulation.
In our model analysis 
we can sum up them by knowing the partition function values
$\left < Z \right >_\sigma$ precisely, but in Monte Carlo simulations
we compute only the average of the observables not the 
partition function.
It is, therefore, extremely important to devise the efficient way 
to perform the multi-thimble integration by extending the Monte Carlo
algorithm for practical applications of the Lefschetz thimble
integration to fermionic systems with the sign problem.

\acknowledgments
H.F.\ was partially supported by JSPS KAKENHI (\# 24540255).
S.K.\ was supported by the Advanced Science Measurement Research Center at Rikkyo University.
Y.K.\ was supported in part by JSPS KAKENHI (\# 24540253).

\appendix

\section{Exact expression and asymptotics of $Z$\label{app:exact}} 
In this appendix, we give the exact expression for the partition
function of the Thirring model with the compact action.
We assume ${\cal N}_{f}=1$ and $L$ is even.

A useful formula for a matrix determinant is known in \cite{Mol:2008}:
\begin{align}
 \det 
\begin{bmatrix}
 a_{1} & b_{1} & 0 & c_{0} \\
 c_{1} & \ddots & \ddots & 0\\
 0  & \ddots & \ddots & b_{L-1}\\
 b_{L} &0 & c_{L-1} & a_{L}
\end{bmatrix} 
 =& -(b_{L} \cdots b_{1} + c_{L-1} \cdots c_{0}) \notag \\
  & + \mathrm{tr} 
\begin{bmatrix}
\begin{pmatrix}
 a_{L} & -b_{L-1} c_{L-1} \\
 1 & 0  
\end{pmatrix} \cdots
\begin{pmatrix}
 a_{2} & -b_{1} c_{1} \\
 1 & 0  
\end{pmatrix} 
\begin{pmatrix}
 a_{1} & -b_{L} c_{0} \\
 1 & 0  
\end{pmatrix} 
\end{bmatrix}. \label{eq:det_form}
\end{align}
In application of this formula to the Dirac operator $D$, the
components $a_{n}, b_{n}$ and $c_{n}$ read 
\begin{align}
& a_{1} = \cdots = a_{L}=m, \notag\\
& b_{n} =  
\begin{cases}
\frac{1}{2}  \e^{+\hat \mu} U_{n-1} & \mbox{for} \ n<L \\
-\frac{1}{2} \e^{+\hat \mu} U_{L-1} & \mbox{for} \ n=L
\end{cases}, \notag \\
& c_{n} = 
\begin{cases}
- \frac{1}{2} \e^{-\hat \mu} U^{-1}_{n-1} & \mbox{for} \ n>0 \\
  \frac{1}{2} \e^{-\hat \mu} U^{-1}_{L-1} & \mbox{for} \  n=0
\end{cases}
\; 
\end{align}
with $U_n=\e^{iA_n}$ and $U_n^{-1}=\e^{-iA_n}$.
Then the 2-by-2 matrix under the trace turns out to be an $L$-th
power of a constant matrix
$\tiny
\left (\begin{array}{cc} 
m & \tfrac{1}{4} \\ 
1 & 0 
\end{array} \right )$.
Now it is straightforward to reach the expression
\begin{align}
 \det D[A] 
&= \frac{1}{2^{L}} \big  [
   2 \cosh \big ( L \hat \mu + i \sum_{n=0}^{L-1} A_{n} \big ) +  m_{+}^{L}
   + m_{-}^{L}  \big ]
\,
\end{align}
where $m_{\pm} = m \pm \sqrt{ m^2+1 }$. 
With $\hat m \equiv \sinh^{-1} m$ and with even $L$,
this can be written as in Eq.~(\ref{eq:detD-A}).

Because the $A_n$-odd terms in the determinant vanish 
after the integration over $A_n$
with weight $\e^{-\beta (1-\cos A_n)}$, 
we can write the partition function as
\begin{align}
Z &=& \frac{1}{2^{L-1}} \int_{-\pi}^{\pi} 
\prod_{n=0}^{L-1} \frac{d A_{n}}{2 \pi} \,
\Big [   \cosh {L\hat \mu}    \prod_{n=0}^{L-1} \cos A_{n} 
+  \cosh {L \hat m}   \Big]
\, \exp \Big ( -\beta \sum_{n=0}^{L-1}( 1-\cos A_{n} )\Big)
\, 
.
\label{eq:part_eff}
\end{align}
This integration is easily performed to yield 
\begin{align}
Z = \frac{\e^{-L \beta} }{2^{L-1}} 
\left[ 
I_{1}(\beta)^{L} \, \cosh {L\hat \mu}  + I_{0}(\beta)^{L}\, \cosh{L \hat m} 
\right],
\end{align}
where $I_{0}(x)$ and $I_{1}(x)$, respectively, are 
the zeroth and first order modified Bessel functions of the first kind.
The fermion number density and the scalar density can be derived by
differentiating $\ln Z$ with respect to $\mu$ and $m$, respectively.


Using the asymptotic expression of 
the modified Bessel function $I_{0,1}(\beta)$ for a large $\beta$,
we find that in the continuum limit at finite $T$,
the partition function
Eq.~(\ref{eq:Zexact}) scales as
\begin{align}
Z &\to \frac{1}{2^{L-1}} \left ( \frac{1}{2\pi \beta } \right )^{L/2}
\e^{-\frac{3g^2}{4T} }
\Big [ \cosh \frac{\mu}{T} +\e^{\frac{g^2}{T}} \cosh \frac{m}{T}\,
\Big ]
,
\end{align}
where we have used $L /\beta =2g^2 /T$ and $L\mu=\mu/T$.
For the uniform-field model (\ref{eq:Z0}),
applying the asymptotic form for large $L$,
\begin{align}
I_L(L\beta) \to \frac{\e^{L \eta}}{\sqrt{2\pi L\,}\, (1+\beta^2)^{1/4}}
\end{align}
with $\eta = (1+\beta^2)^{1/2} + \log
\frac{\beta}{1+(1+\beta^2)^{1/2}}$,
we find
\begin{align}
Z_0 & \to 
\frac{1}{2^{L-1}} \left ( \frac{1}{2\pi L \beta } \right )^{1/2}
\e^{-\frac{g^2}{T} }
\Big [ \cosh \frac{\mu}{T} +\e^{\frac{g^2}{T}} \cosh \frac{m}{T} \,
\Big ]
\, .
\end{align}
It is interesting to observe that 
in the $T \to 0$ limit both models 
show a first order transition at the same point $|\mu_c|=m+g^2$.

If we take $L$ large with $\beta$ fixed,
we find
\begin{align}
Z & \to \frac{1}{2^L} \left ( \frac{1}{2\pi \beta } \right )^{L/2}
\Big [ I_1(\beta)^L \e^{L |\hat \mu|} +I_0(\beta)^L \e^{L \hat m}
\Big ]
,
\end{align}
and for the uniform-field model
\begin{align}
Z_0 & \to 
\frac{1}{2^L} \left ( \frac{1}{2\pi L\beta } \right )^{1/2}
\left [
\frac{\sqrt{\beta\,}\, \e^{L(\eta -\beta)}} {(1+\beta^2)^{1/4}}
\e^{L |\hat \mu|}  + \e^{L \hat m}
\right ]
\, .
\end{align}
In the infinite-$L$ limit these models show 
a first-order transition
at $|\hat \mu_c| = \hat m + \ln (I_0(\beta)/I_1(\beta))$
and $|\hat \mu_c| = \hat m + \beta - \eta$, respectively.



\end{document}